\documentclass[11pt,a4paper]{article}
\pdfoutput=1
\usepackage{jheppub}
\usepackage{bm}
\usepackage{feynmf}
\usepackage{mathrsfs}
\usepackage{slashed}
\usepackage{xcolor}
\usepackage{rotating}
\usepackage{transparent}
\usepackage{amsmath}
\usepackage{orcidlink}

\allowdisplaybreaks

\newcommand{\real}{\mathbb{R}}
\newcommand{\complex}{\mathbb{C}}
\newcommand{\del}{\partial}
\newcommand{\ud}{\mathrm{d}}
\newcommand{\I}{\mathcal{I}}

\newcommand{\cl}{\mathrm{cl}}
\newcommand{\q}{\mathrm{q}}

\title{Computing real-time quantum path integrals on Sewed, almost-Lefschetz thimbles}
\author[a,b]{Zong-Gang Mou\orcidlink{0000-0002-0986-2366},}
\author[b]{Paul M. Saffin\orcidlink{0000-0002-4290-3377},}
\author[c]{Anders Tranberg\orcidlink{0009-0001-4033-1395}}
\date{~}

\affiliation[a]{School of Physics and Astronomy, University of Southampton,  Highfield, Southampton, SO17 1BJ, United Kingdom,}
\affiliation[b]{School of Physics and Astronomy, University of Nottingham, University Park, Nottingham NG7
2RD, United Kingdom,}
\affiliation[c]{Faculty of Science and Technology, University of Stavanger, 4036 Stavanger, Norway}

\emailAdd{z.g.mou@soton.ac.uk}
\emailAdd{paul.saffin@nottingham.ac.uk}
\emailAdd{anders.tranberg@uis.no}

\abstract{
 We present a method to compute real-time path integrals numerically, by Monte-Carlo sampling on near-Lefschetz thimbles. We present a collection of tools based on the Lefschetz thimble methods, which together provide an alternative to existing methods such as the Generalised thimble. These involve a convenient coordinate parameterization of the thimble, direct numerical integration along a radial coordinate into an effective path integral weight and locally deforming the Lefschetz thimble using its Gaussian (non-interacting theory) counterpart in a region about the critical point. We apply this to quantum mechanics, identify possible pitfalls and benefits, and benchmark its efficiency. 
}

\begin{document} 
\maketitle
\flushbottom

%%%%%%%%%%%%%%%%%%%%%%%%%%%%%%%%%%%%%%%%%%%%%%%%%%%%%
\section{Introduction}
%%%%%%%%%%%%%%%%%%%%%%%%%%%%%%%%

In Quantum Field Theory (QFT), we usually seek to compute the expectation value of operators ${\mathcal O}(\hat\phi)$, for some general function of the basic field operator $\hat\phi$, and we can do that using a path integral over some probability distribution function $e^{-\I}$, such as,
\begin{align}
\langle {\mathcal O}\rangle \equiv \frac{\int{\mathcal D}\phi \;e^{-\I} {\mathcal O}}{\int {\mathcal D}\phi\;  e^{-\I}}.
\end{align}
The integrals are over real-valued fields $\phi$, and the interval of integration ranges from $-\infty$ to $\infty$. 

If the exponent $\mathcal{I}$ of the weight is real-valued, such integrals may be computed analytically or through numerical Monte-Carlo sampling \cite{Creutz:1983njd}. In QFT this may be achieved directly for systems at finite temperature  or for some complex exponents by Wick rotation to Euclidean time \cite{Laine:2016hma}. 

For truly real-time evolution out of equilibrium, Wick rotation is not applicable, $\mathcal{I}$ is explicitly imaginary, the exponential is a complex phase, and straightforward convergence of the integral is lost. This is known as the ``sign" problem \cite{Alexandru:2020wrj}, and represents a central problem of contemporary QFT.

Several approaches have been suggested, and are still under development to resolve or at least ameliorate this sign problem. Two of these involve allowing the field variables to take on complex values either through sampling by a Langevin equation exploring the entire complex plane (complex stochastic quantization \cite{Aarts:2008rr,Seiler:2012wz,Aarts:2016qrv}), or restricting Monte-Carlo sampling to a deformation of the real axis into the complex plane (or $\mathbb{R}^n$ into $\mathbb{C}^n$) (Lefschetz/Generalised Thimbles \cite{Witten:2010cx,Cristoforetti:2013wha,Cristoforetti:2012su,DiRenzo:2015foa,Alexandru:2016gsd, Alexandru:2017lqr,Mou:2019tck,Mou:2019gyl,DiRenzo:2021kcw,Nishimura:2023dky, Nishimura:2024bou}). 

The former has been very successful in describing simple systems, and while a few foundational issues remain unresolved, it has now also been tamed enough to be applied to equilibrium QCD at finite density \cite{Aarts:2013uxa,Sexty:2013ica,Ito:2020mys}. The latter is based on deep mathematics \cite{Witten:2010cx}, but has been found to be numerically demanding. The approach requires more work to allow scaling to physically relevant system sizes \cite{Alexandru:2020wrj,Nishimura:2024bou}.

A subtle difference among different thimble approaches can also be readily reviewed.
A Lefschetz thimble in high dimension can only be determined numerically by the flow equation from its critical point.
Since the critical point is a fixed point, the flow equation starts in practice from a small region around the critical point where the Gaussian approximation is good enough.
The constructed surface becomes closer to the Lefschetz thimble when the small region gets closer to the critical point \cite{Cristoforetti:2013wha,Cristoforetti:2012su,DiRenzo:2015foa}.
In comparison, the Generalized Thimble is exact from the very beginning.
Any integral in the family is supposed to attain the same result as the original integral yields, thanks to Cauchy's integral theorem, and for infinite flow time the generalized thimble will reach the appropriate set of Lefschetz thimbles
\cite{Alexandru:2016gsd, Alexandru:2017lqr}.
In light of the Generalized Thimble/Cauchy's integral theorem, we reexamine the Lefschetz thimble and introduce another family of integration cycles, {\it sewed thimbles}. The integration over these surfaces exactly reproduces the required result and the Lefschetz thimble will be recovered in a particular limit.

In the present work, we present a set of developments to facilitate Monte-Carlo sampling (almost) directly on a Lefschetz thimble. We have in mind the situation of an initial-value problem, where initial conditions (the field and its derivative) are sampled by other means \cite{Mou:2019tck,Mou:2019gyl}. The subsequent quantum time-evolution of observables can then be determined through the use of thimble methods. In particular, by considering one member of an ensemble of initial conditions at a time, there is no issue of having to sample over multiple thimbles. Each initial condition corresponds to a unique thimble. 

The paper is organised as follows: We first introduce our method for a toy model double integral in section \ref{sec:toymodel}. We introduce the ``Sewed" thimble, polar coordinates, stereographic coordinates  and the integration over rays. In section \ref{sec:discretized} we set up the same formalism for real-time QFT of a single scalar field in 0+1 dimensions and carry out the computation on small systems. We conclude in section \ref{sec:conclusion}.

%%%%%%%%%%%%%%%%%%%%%%%%%%%%
\section{Setting up the formalism in 2 dimensions}
\label{sec:toymodel}
%%%%%%%%%%%%%%%%%%%%%%%%%%%

\begin{figure}[t]
\centering
\includegraphics[width=0.7\linewidth,clip]{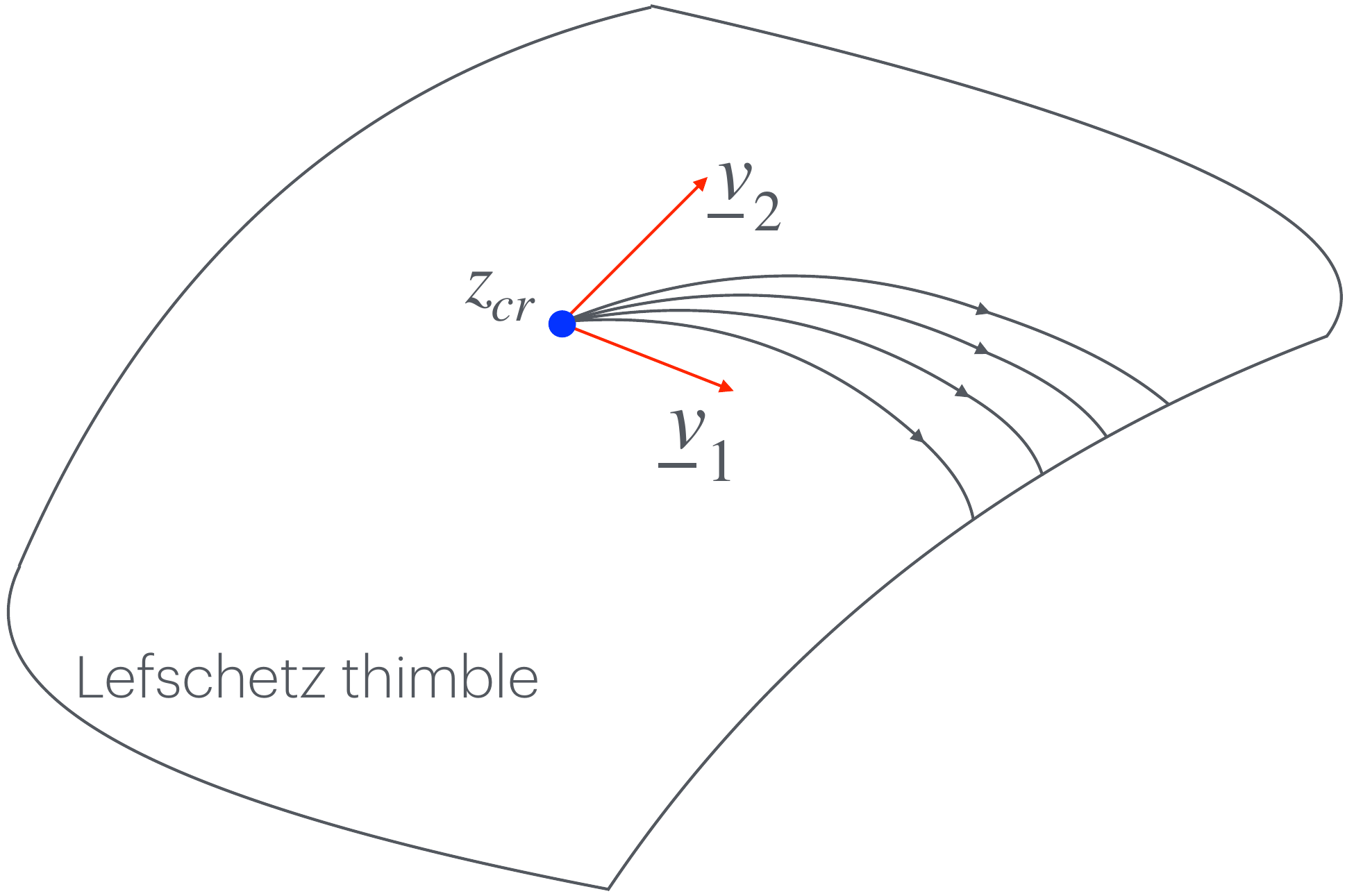} 
\caption{A schematic of the Lefschetz thimble embedded in $\complex^2$. The figure shows the gradient flow curves emanating from the critical point, that together make up the Lefschetz thimble, along with eigenvectors of the Hessian of $\I$, $\underline v_\alpha$.}
\label{fig:lefschetz_thimble}
\end{figure}

As a starting point for our discussion, we consider a proxy for the sort of problems we are going to face in the quantum mechanical path integral, and consider the ratio of two-dimensional integrals
\begin{align}\label{eq:O_expec}
    \langle\mathcal{O}\rangle&=\frac{\int_{\real^2} \ud x\,\ud y \; \mathcal{O}(x,y)\;e^{-\I}}{\int_{\real^2} \ud x\,\ud y \;e^{-\I}},
\end{align}
where $\I$ is, in general, a complex-valued function. The procedure one may follow to perform these integrals is to promote $x$ and $y$ to complex  variables, 
\begin{align}
    (x,y)\to(z_1,z_2),
\end{align}
and deform the integration manifold from $real(z_1)\times real(z_2)$ to some more suitable manifold (comprising the thimbles) living in the $\complex^2$ parametrized by the $\underline z$. The question then arises as to how to construct this ``suitable" manifold. 
One such choice is
the Lefschetz thimbles, manifolds associated with
critical (stationary) points of $\I(\underline z)$ in $\complex^2$, with one thimble for each critical point. The Lefschetz thimble is defined as the collection of gradient flow curves that emanate from the critical point\footnote{Note that the flow is zero at the critical point, so we are really interested in those curves that solve the flow equation, with the condition that $\underline z(\tau\to-\infty)=\underline z_{cr}$}, with the flow determined through\footnote{We will take overbar and $^*$ to denote complex/hermitian conjugation.}
\begin{align}\label{eq:flow_equation}
\frac{\ud z_i}{\ud \tau} = \overline{\frac{\partial \I}{\partial z_i}}.
\end{align}
This process is sketched
in Figure \ref{fig:lefschetz_thimble}.
Also shown in the figure is the set of eigenvectors of the Hessian of $\I$, which we shall find useful later for parametrizing the thimble. In some systems there may be multiple critical points, but not all of the thimbles will contribute to the integral. To determine which do, one considers the {\it upwards} gradient flow, i.e. (\ref{eq:flow_equation}) with a minus sign. The critical points whose {\it upward} flows intersect the original integration manifold $\real^2$ are the ones whose thimbles contribute to the integral \cite{Witten:2010cx}. In our systems we shall only have a single critical point, and so this subtlety does not arise \cite{Mou:2019tck,Mou:2019gyl}.

Starting from any point, other than $z_{cr}$, the gradient flow leads to increasing ${\rm Re}(\I)$, while keeping the imaginary part, ${\rm Im}(\I)$, constant. This follows straightforwardly from the flow equation, 
\begin{align}
\frac{\ud \I}{\ud \tau}= \frac{\partial \I}{\partial z_i}\frac{\ud z_i}{\ud \tau} = \left|\frac{\partial \I}{\partial z_i}\right|^2.
\end{align}
As a result, the whole integrand $e^{-\I}$ becomes exponentially suppressed away from the critical point along the flow. In effect, whereas the integral over the original manifold picks up contributions from the entire real axis, only the region near the critical point contributes when integrating over the complex-valued thimble. This property 
is what makes the integral tractable numerically.

An alternative procedure is to use the same flow equation to flow all the points on the real manifold $\real^2$ into $\mathbb{C}^2$. Then (the 2-dimensional generalisation of) Cauchy's integral theorem applies straightforwardly to an integral over this continuous deformation of the real space, a {\it Generalised thimble} \cite{Alexandru:2016gsd,Alexandru:2017lqr}. The flow equation again ensures that the integral becomes suppressed as the flow proceeds, and the trick is then to find a suitable flow time $\tau$ where the integral converges sufficiently fast, whilst also keeping the Generalised thimble smooth enough to efficiently Monte-Carlo sample it.

\begin{figure}[t]
\centering
\includegraphics[width=0.7\linewidth,clip]{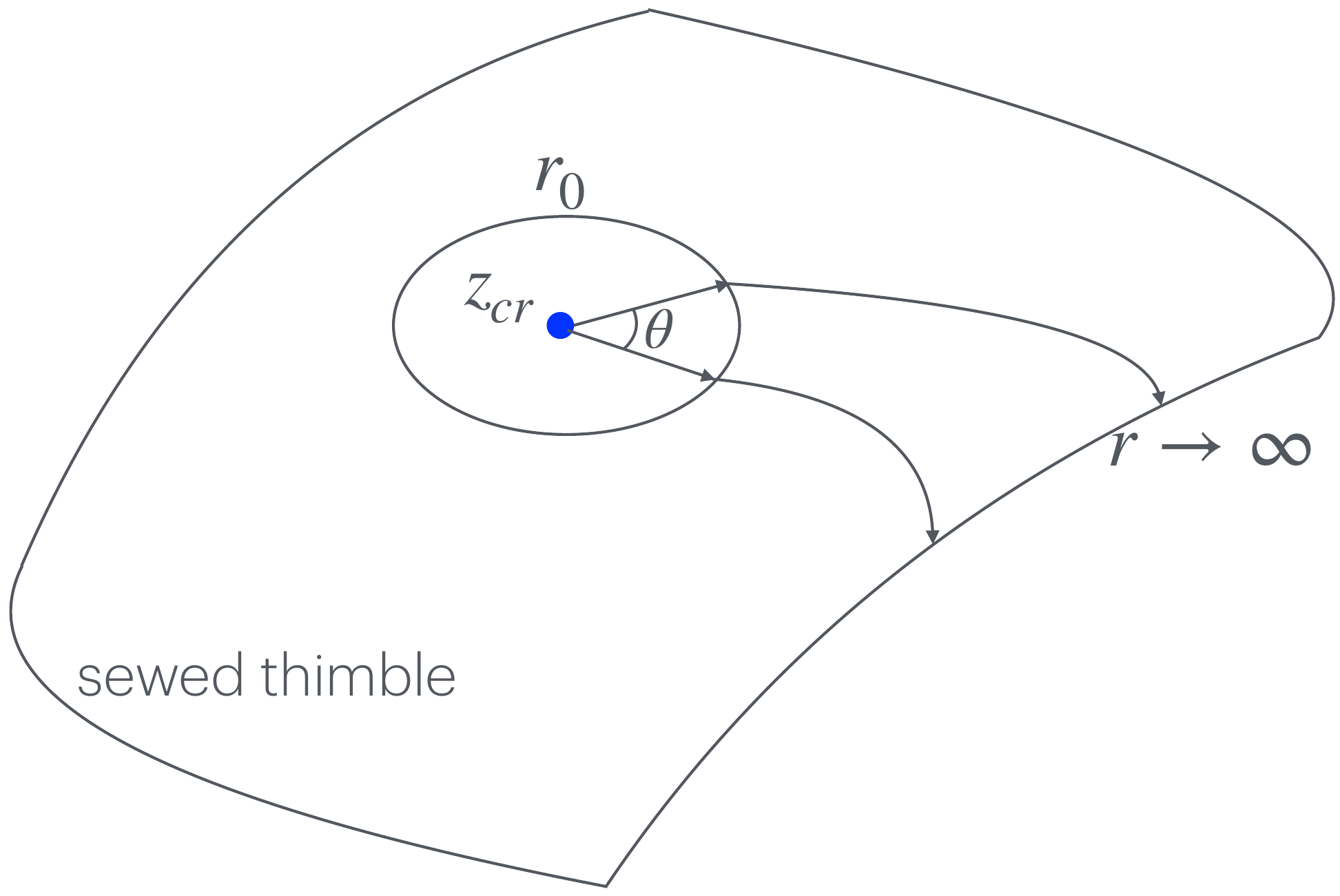} 
\caption{A schematic of the construction of a sewed thimble. The sewed thimble is the curve that starts at the critical point, using the quadratic part of $\I$ to flow until the curve reaches some distance $r_0$ from the critical point. At this point the full exponent $\I$ is used to construct the flow for the remainder of the curve. }
\label{fig:sewed_thimble}
\end{figure}

The two, related, proposals in this paper are to use a modified version of the Lefschetz thimble, and to parametrize the thimble in a natural polar way that is suggested by the system itself. The {\it Sewed thimble} that we propose comprises two regions, the inner region is a surface constructed from the gradient flow of the quadratic part of $\I$, while the outer is constructed as the gradient flow of the full $\I$. The size of this inner region is a matter of choice, depending on the system. It is important to recognise that this does not constitute an approximation to the integral, as Cauchy's theorem still applies to this sewed thimble, rather it is an approximation to the Lefschetz thimble. The parametrization that we use to describe the thimble is polar in nature, with the flow time providing a radial co-ordinate, $r\sim e^\tau$, and the different rays (flows) being determined by angular co-ordinates, as depicted in Figure \ref{fig:sewed_thimble}. The aim is then to solve the integral (\ref{eq:O_expec}) by sampling in the angular directions, while performing the full radial integral of the flow for each instance of the angles. 

%%%%%%%%%%%%%%%%%%%%
\subsection{Flow, inner and outer regions}
%%%%%%%%%%%%%%%%%%
We start by examining the inner region of the sewed thimble, where we use the quadratic expansion of $\I$ about the critical point,
\begin{align}
    \I_{inner}&=\frac{1}{2}z_iH_{ij}z_j,
\end{align}
where
\begin{align}
    H_{ij}&=\left.\frac{\del^2\I}{\del z_i\del z_j}\right |_{z_{cr}},
\end{align}
such that the flow equation in the inner region is
\begin{align}
    \frac{\ud z_i}{\ud\tau}&=(H_{ij}z_j)^\star.
\end{align}
In these expressions the indices range over $x$ and $y$ for our example, but more generally they range over the dimension of the integration manifold, $1,...,n$. We have also made the co-ordinate choice that the critical point is at $\underline z_{cr}=0$. Given the linearity of this flow equation, we are able to solve it analytically \cite{Mukherjee:2013aga,Fujii:2013sra}. However, rather than trying to minimize the size of this inner region to end up at the Lefschetz thimble, we embrace it, and let its size become a variable that may be tuned. 

To proceed we make use of Takagi factorization \cite{Horn_Johnson_2012,Fujii:2013sra} and write
\begin{align}
    v^THv&=D,
\end{align}
where $v$ is a unitary matrix, and $D$ is a diagonal matrix whose entries are the positive square roots of the eigenvalues of $H^\dagger H$. This leads us to an eigenvalue problem,
\begin{align}
    Hv&=(Dv)^\star,\\\label{eq:H_eigen}
    \Rightarrow H_{ij}v_{j\alpha}&=\kappa_\alpha v_{i\alpha}^\star\qquad\textnormal{no sum on $\alpha$.}
\end{align}
We interpret the $\alpha$ index as labelling the eigenvector $\underline v_\alpha$\footnote{Note that the right hand side of (\ref{eq:H_eigen}) involves $\underline v_\alpha^\star$, rather than $\underline v_\alpha$, so this is not the usual eigenvalue/eigenvector problem.} of $H$.

We now parametrize the thimble by introducing the real co-ordinates $\xi^\alpha$, such that on the thimble
\begin{align}\label{eq:z_v_xi}
    z_i&=\sum_\alpha v_{i\alpha}\xi_\alpha,
\end{align}
and we discover that the flow equation may now be solved to find
\begin{align}
    \xi_\alpha(\tau)&=\tilde e_\alpha \exp(\kappa\tau),
\end{align}
and we then trade one of the $\tilde e_\alpha$ integration constants for a constant, $\tau_{\rm shift}$, by writing
\begin{align}\label{eq:xi_sol}
    \xi_\alpha(\tau)&=e_\alpha\exp[\kappa_\alpha(\tau+\tau_{\rm shift})],\\\label{eq:e_constraint}
    \sum_\alpha e_\alpha e_\alpha&=1.
\end{align}
This decomposition is just a polar co-ordinate description of the surface, with the $e_\alpha$ describing the angular co-ordinates on the thimble giving the direction of the flow, and the $e^\tau$ being the radial co-ordinate.
Since the flow equation is autonomous, there exists a time shift symmetry in the solutions, and we can simply fix $\tau_{\rm shift}$ to some constant without loss of generality.
It is important to understand the co-ordinates on the thimble because any integration on the thimble is going to require Jabobian factors, meaning we need an expression for the components of $\frac{\del z_i}{\del\tau}$ and $\frac{\del z_i}{\del e_\alpha}$. $\frac{\del z_i}{\del\tau}$ is just the flow equation (\ref{eq:flow_equation}), because the $e_\alpha$ just label which flow we are on, and so are constant along a given flow. 
On the other hand, $\frac{\del z_i}{\del e_\alpha}$ can simply be obtained from differentiation of equation (\ref{eq:z_v_xi}).
We now have enough information to compute the change in the integration measure
\begin{align}\label{eq:coord_trans_measure}
    \ud^n z&=\mathrm{det}\left( \frac{\del (z_1,z_2,..)}{\del(\tau,e_1,e_2,...)}\right)\ud\tau\,\ud e^1\,\ud e^2...
\end{align}
Having flowed up to some distance $e^\tau=r_0$ we then use the full non-linear $\I$ to construct the remainder of the flow. The co-ordinates remain $\tau$ ($~$the distance along the flow) and $e_\alpha$ (the label of the flow), and so we need to know how to compute the Jacobian in this outer region. Again, the $\frac{\del z_i}{\del\tau}$ of the Jacobian just comes from the flow equation (\ref{eq:flow_equation}), but now with the full $\I$, while for $\frac{\del z_i}{\del e_\alpha}$ we note that the flow equation leads to
\begin{align}
\frac{\del}{\del \tau}\left(\frac{\partial z_i}{\partial e_\alpha}\right) =\overline{\frac{\partial^2\I}{\partial z_i\partial z_j}\frac{\partial z_j}{\partial e_\alpha}},
\label{eq:outerint}
\end{align}
which means we may take the expression for $\frac{\partial z_i}{\partial e_\alpha}$ at the boundary of the inner region, and use (\ref{eq:outerint}) to propagate it along the flow, giving the remaining parts of the Jacobian of (\ref{eq:coord_trans_measure}).

By varying the size of the inner region through the parameter $r_0$ ($\tau_0$), we obtain a parametric family of integration manifolds. When $r_0\to 0$, the Sewed thimble converges to the Lefschetz thimble.
Thanks to the Cauchy's integral theorem, the result of the integral on the new integration contour will be the same as on the Lefschetz thimble for any $r_0$.

%%%%%%%%%%%%%%%%%%%%%%%%%%%%%%%%%
\subsection{Application to a toy model}
%%%%%%%%%%%%%%%%%%%%%%%%%%%%%%%%%%

With the outline of the formalism given, let us work through an example where we have analytic control, and consider the following expectation value
\begin{align}
\langle y^4\rangle = \frac{\int \ud x\,\ud y \; y^4\;e^{-iax(y-b)-icx^3}}{\int \ud x\,\ud y \;e^{-iax(y-b)-icx^3}},
\label{eq:1variable}
\end{align}
where without loss of generality we set $a>0$. In this simple example, we can obtain the analytic results (with a detailed derivation in Appendix \ref{app:B})
\begin{align}
&\int \ud x\,\ud y \;e^{-iax(y-b)-icx^3} = \frac{2\pi}{a}
,\\
&\int \ud x\,\ud y \;y^4\;e^{-iax(y-b)-icx^3} = \frac{2\pi}{a}\left(b^4+\frac{24bc}{a^3}\right).
\end{align}
In this context, the factor $y^4$ in the numerator's integrand plays the role of the observable $\mathcal{O}$, while the exponential function is the weight $e^{-\mathcal{I}}$.
For more complicated operators, % such as $\langle x^2y^4\rangle$, 
the analytical result is not as readily available, and we turn to Lefschetz thimble methods to compute the integral.

To compute the Lefschetz thimble one first promotes the real variables to complex ones, 
\begin{align}
    (x,y)\to(\tilde z_1,\tilde z_2)\quad
    \Rightarrow \I=ia\tilde z_1(\tilde z_2-b)+ic\tilde z_1^3,
\end{align}
and find the critical point(s) of $\I$. In this case there is a single critical point given by
\begin{align}\label{eq:general_flow_eqn}
    \underline {\tilde z}_{cr}=(0,b)^T.
\end{align}
The Lefschetz thimble then follows from constructing all the downward gradient flows (\ref{eq:flow_equation}) that come from the critical point as $\tau\to-\infty$. It is useful to adapt the co-ordinates to those centred about the critical point, giving
\begin{align}
    \underline{z}&=\underline{\tilde z}-\underline{\tilde z}_{cr},\\\label{eq:I_z_coords}
    \I&=iaz_1z_2+icz_1^3.
\end{align}

%%%%%%%%%%%%%%%%%%%%%%%
\subsubsection*{Inner integral:}
%%%%%%%%%%%%%%%%%%%%%%%%%%

The Gaussian thimble admits an analytic coordinate description, which we will use for the inner region.
To be precise, the Gaussian thimble is the Lefschetz thimble for the action up to the quadratic terms around the critical point,
\begin{align}\label{eq:IG}
\I_G=\I_0 + \frac{1}{2}\underline z^T H \underline z,
\end{align}
where in our toy model (\ref{eq:I_z_coords}), $\I_0=0$, and
\begin{align}
H= \left( \begin{array}{cc} 0 &  ia \\ ia & 0 \end{array} \right).
\end{align}
the (positive) eigenvalues and eigenvectors of $H$, 
$H\underline v=\kappa\underline{v}^\star$,
are found to be
\begin{align}
\kappa_1=a
,\quad
\underline v_1=\frac{1}{\sqrt{2}}e^{-i\pi/4}
\left(
\begin{array}{l}
1 \\ 1
\end{array}
\right)
,\quad\quad
\kappa_2=a
,\quad
\underline v_2=\frac{1}{\sqrt{2}}e^{i\pi/4}
\left(
\begin{array}{c}
-1 \\ 1
\end{array}
\right).
\end{align}
Then (\ref{eq:xi_sol}), (\ref{eq:z_v_xi}) leads to
\begin{align}
    z_1&=\frac{1}{\sqrt{2}}e^{-i\pi/4}p_0r^ae_1-\frac{1}{\sqrt{2}}e^{i\pi/4}p_0r^ae_2,\\
    z_2&=\frac{1}{\sqrt{2}}e^{-i\pi/4}p_0r^ae_1+\frac{1}{\sqrt{2}}e^{i\pi/4}p_0r^ae_2,\\
    r&=e^\tau,\qquad p_0=e^{a\tau_{\rm shift}},
\end{align}
which do indeed satisfy the flow equation (\ref{eq:flow_equation}) for the quadratic part of $\I$. Noting the constraint (\ref{eq:e_constraint}) we choose
\begin{align}\label{eq:polar_co-ords}
    e_1=\cos\theta,\qquad e_2=\sin\theta,
\end{align}
to find that the $\underline z$ co-ordinates of the inner region of the sewed thimble, parametrized by radial co-ordinate $r$ and angular co-ordinate $\theta$, are
\begin{align}
\begin{array}{l}
z_1=\frac{1}{\sqrt{2}}p_0r^ae^{-i\theta}e^{-i\pi/4},
\\
z_2=\frac{1}{\sqrt{2}}p_0r^ae^{i\theta}e^{-i\pi/4}.
\end{array}
\end{align}
In order to perform the integrals in the new radial and angular co-ordinates we need the Jacobian,
\begin{align}
J=\frac{\partial (z_1,z_2)}{\partial (r,\theta)}
=
\frac{1}{\sqrt{2}}p_0e^{-i\pi/4}
\left(
\begin{array}{cc}
ar^{a-1}e^{-i\theta} & -ir^ae^{-i\theta}
\\
ar^{a-1}e^{i\theta} & ir^ae^{i\theta}
\end{array}
\right),
\end{align}
and the numerator integral of (\ref{eq:1variable}) in $(r,\theta)$-coordinates for the inner part of the sewed thimble takes the form
\begin{align}
\int_{0}^{2\pi}\ud\theta \int_{0}^{r_0}\ud r \;ap_0^2r^{2a-1}
\exp\left[
-\frac{1}{2}ap_0^2r^{2a}-ic\left(\frac{p_0}{\sqrt{2}}r^ae^{-i\theta}e^{-i\pi/4}\right)^3
\right]
{\mathcal O}[x(r,\theta),y(r,\theta)],
\label{eq:innerint}
\end{align}
where we consider a general function ${\mathcal O}$, and not just $y^4$.

%%%%%%%%%%%%%%%%%%%%%%%
\subsubsection*{Outer integral:}
%%%%%%%%%%%%%%%%%%%%%%

Having found the relevant integral for the inner part of the sewed thimble we now need to compute the contribution from the outer part. The gradient flow equation with the full $\I$ defines the sewed thimble in the outer region, starting from the flow time $\tau_0\equiv \ln(r_0)$ and flowing to infinity, which is carried out numerically. The integral on the thimble is computed through
\begin{align}\label{eq:theta_r_integral}
\int_{0}^{2\pi}\ud\theta \int_{\tau_0}^{+\infty}\ud \tau \;{\rm Det}(J) e^{-\I}
{\mathcal O}(x,y),
\end{align}
where the Jacobian matrix is given by
\begin{align}
J= \frac{\partial (z_1,z_2)}{\partial (\tau,\theta)}
=\left(\frac{\del \underline z}{\partial \tau},\frac{\partial \underline z}{\partial\theta}\right).
\end{align}
The $\frac{\del \underline z}{\partial \tau}$ part of the Jacobian comes from the flow equation itself, as we know $\frac{\del\I}{\del\underline z}$ once the location on the thimble, $\underline z$, is given. For the $\frac{\partial \underline z}{\partial \theta}$ part of the Jacobian, we know its value at $r_0$ because we have the analytic expression for $\underline z$ in the inner region of the sewed thimble. We then take the value of $\frac{\partial \underline z}{\partial \theta}$ at $r_0$ and find its value along the flow by solving (\ref{eq:outerint}), expressed in $\theta$
\begin{align}
\frac{\partial}{\partial \tau}\left(\frac{\partial z_i}{\partial \theta}\right) =\overline{\frac{\partial^2\I}{\partial z_i\partial z_j}\frac{\partial z_j}{\partial\theta}}.
\end{align}
The initial values for the flow starting at $r_0$ are
\begin{align}
\tilde z_1(r_0)=\frac{1}{\sqrt{2}}p_0r_0^ae^{-i\theta}e^{-i\pi/4}, \quad
\tilde z_2(r_0)=b+\frac{1}{\sqrt{2}}p_0r_0^ae^{i\theta}e^{-i\pi/4}
,\\
J(r_0)
=
\left(
\begin{array}{cc}
\frac{p_0}{\sqrt{2}}ar_0^{a}e^{-i\theta}e^{-i\pi/4}
+3cr_0^{2a}\left(\frac{p_0}{\sqrt{2}}e^{i\theta}\right)^2
&
-i\frac{1}{\sqrt{2}}p_0r_0^ae^{-i\theta}e^{-i\pi/4}
\\
\frac{p_0}{\sqrt{2}}ar_0^{a}e^{i\theta}e^{-i\pi/4}
&
i\frac{1}{\sqrt{2}}p_0r_0^ae^{i\theta}e^{-i\pi/4}
\end{array}
\right),
\end{align}
where we have reverted to the original co-ordinates $\underline{\tilde z}$, rather than those centred about the critical point.
\begin{figure}[t]
\begin{tabular}{cc}
\includegraphics[width=0.5\linewidth,trim=0.cm 0cm 0cm 0cm, clip]{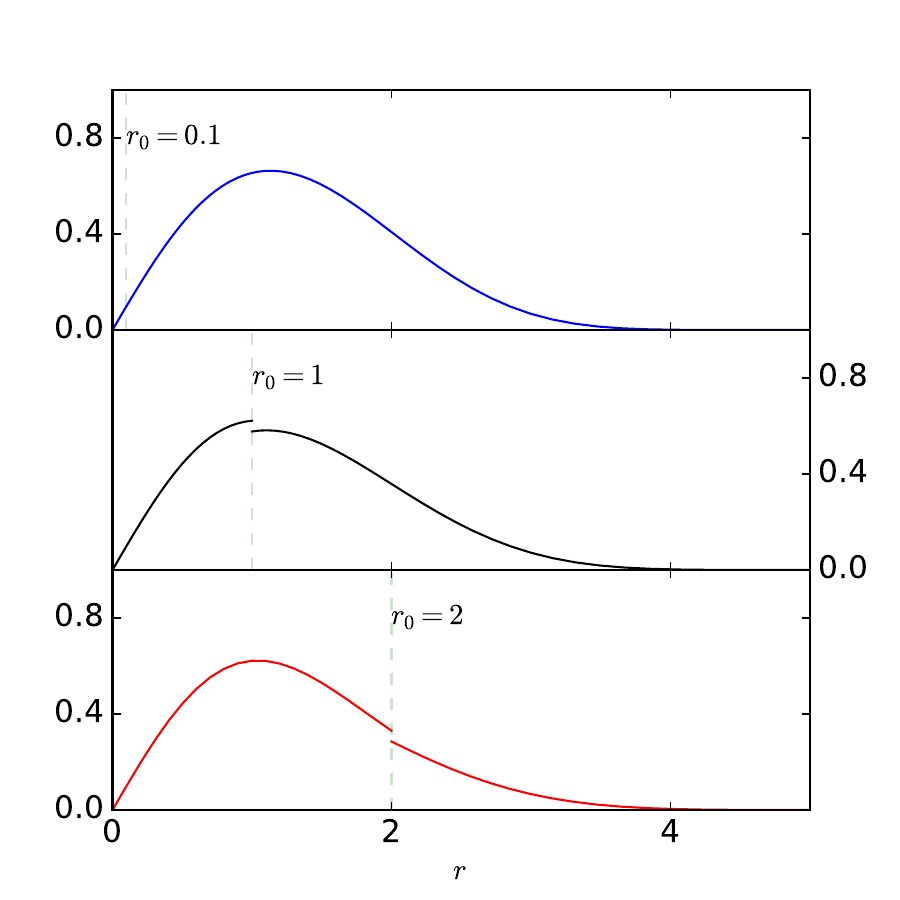} &
\includegraphics[width=0.5\linewidth,trim=0.cm 0cm 0cm 0cm, clip]{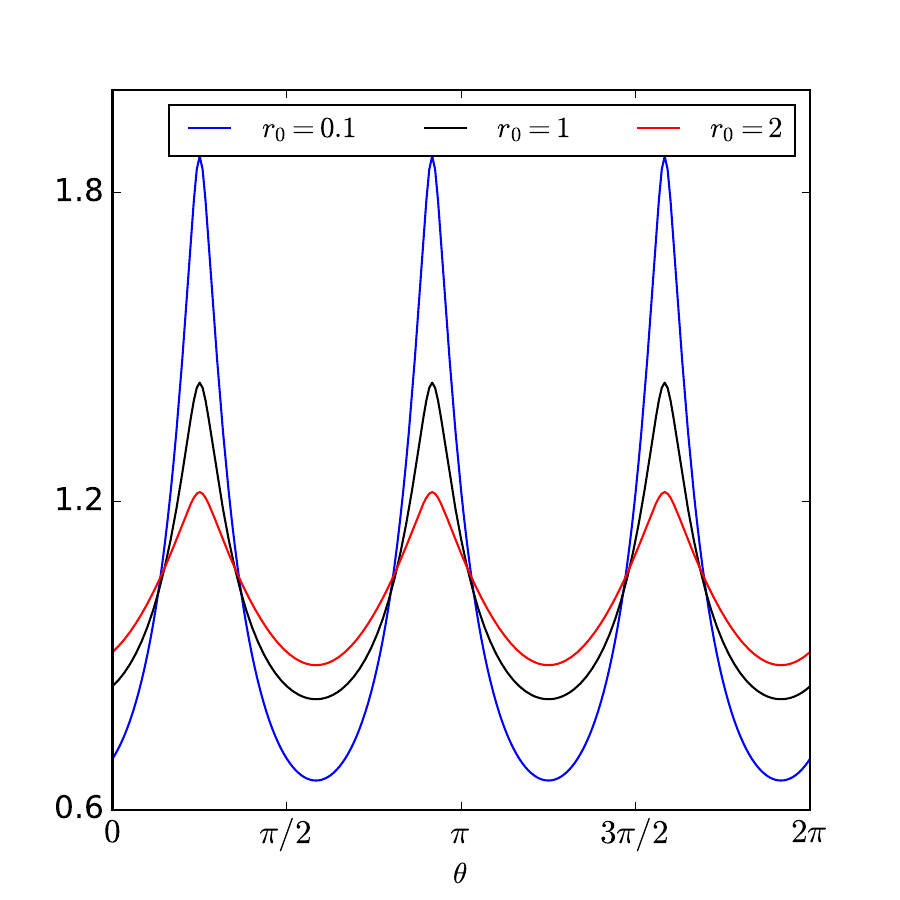}
\end{tabular}
\caption{
$p_0=1$,
$a=1$,
$b=1$,
$c=0.1$.
(L)
The weight functions $\exp\left(-{\rm Re}(\I)+\ln|{\rm Det}J|\right)$, against the flow distance $r$, for a particular choice of $\theta=\pi/3$.
(R) The integrated (over $r$) weight function $\int \ud r \exp\left(-{\rm Re}(\I)+\ln|{\rm Det}J|\right)$, against the angle $\theta$.
}
\label{fig:r_theta}
\end{figure}

In this sewed thimble approach we have a new parameter, $r_0$, which defines the crossover point between flowing with the Gaussian exponent $\I_G$ and the full exponent $\I$. The integration region is parametrized by $r$ and $\theta$, and we can examine how the integrand behaves along a particular $\theta$ direction as we move along the flow parametrized by the radius $r$. 
The integrand of (\ref{eq:theta_r_integral}) naturally splits into two pieces, a real weighting factor, $e^{-\textrm{Re}(\I)+\ln({\rm Det}J)}$, and the rest, which comprises the phase factors $e^{i\left\{{\rm Arg}({\rm Det}J))-\textrm{Im}(\I)\right\}}$ and the $\mathcal{O}$ term. In Figure \ref{fig:r_theta} (left), we show the logarithm of the weight function for the inner (\ref{eq:innerint}) and outer (\ref{eq:theta_r_integral}) parts of the integral, for different values of $r_0$, along a particular direction, $\theta=\pi/3$. The integrand contains the entire action $\I$ on the entire thimble, but the Jacobian matrix in the inner region only includes the Gaussian contribution.
This introduces a discontinuity in the complete integrand, which is unproblematic, as we do the inner and outer integrals separately. We see that the integrand decays as we move along the flow, which is the property we need in order to render the integral well behaved at large $r$. 

With this setup, we are now able to compute the integrals numerically. We can first perform the integral along the flow time $\tau$ (equivalently the flow distance $r$), where exponential suppression allows for convergence for a fairly small range in $\tau$. This radial integral (via fourth-order Runge–Kutta method in practice, over both the inner range $r=0-r_0$ and the outer range $r_0-\infty$) leads to an effective weight function, now only dependent on the angle $\theta$, as shown in Figure \ref{fig:r_theta} (right). We observe that in this toy model example, the weight function is quite dependent on $r_0$.
Although the integrand can have strong $\theta$ and $r_0$-dependence, the integral remains finite and independent of $r_0$. Still, when Monte-Carlo sampling the integrand, one would struggle with ergodicity if the peaks become too sharp. Analogous peaks were seen in the partial partition function of the one-dimensional Thirring model \cite{DiRenzo:2021kcw}.

Finally, we can easily perform the integral numerically along the angular direction, from $0$ to $2\pi$, to the required numerical precision. Despite having different choices of $r_0$ and different weight functions, the total integral will be the same. This direct prescription for the integral is only applicable for low dimensional integrals, with few integration variables. In higher dimensions, we must turn to Monte Carlo integration. To get a feel for how that is done, we will proceed with our toy model. 

%%%%%%%%%%%%%%%%%%%%%%%%%%%%%%%%%%%%%%%%%%
\subsection{Monte Carlo sampling and reweighting}
\label{sec:1variable}
%%%%%%%%%%%%%%%%%%%%%%%%%%%%%%%%%%%%%%%%%%

Continuing with our parametrization of the Sewed thimble from the previous section, we can now instead use Monte Carlo sampling to compute
\begin{align}
\langle {\mathcal O} \rangle
=\frac{
\int \ud\theta\, \ud\tau\; e^{-i{\rm Im}(\I)+i{\rm Arg}({\rm Det}J)} e^{-{\rm Re}(\I)+\ln|{\rm Det}J|} {\mathcal O}(\tau,\theta)
}{
\int \ud\theta\, \ud\tau \;e^{-i{\rm Im}(\I)+i{\rm Arg}({\rm Det}J)} e^{-{\rm Re}(\I)+\ln|{\rm Det}J|}
}.
\label{eq:O_theta}
\end{align}
We have explicitly split the weight function up into a positive definite part and a complex phase. Integrating over the real-valued domain, we encounter the classic sign problem. The integrand $e^{-iax(y-b)-icx^3}$ contributes with equal amplitude $|e^{-\I}|=1$ for all $x,y$, while the complex phase can be arbitrary. Summing over contributions with different phases leads to cancellations, so that the denominator is much smaller than its statistical errors. 

On the Sewed thimble, we can carry out the Monte Carlo sampling methods like on the Lefschetz thimble, either by point $(\tau,\theta)$ \cite{Cristoforetti:2013wha,Cristoforetti:2012su}, or by ray $(\theta)$ \cite{DiRenzo:2015foa}.
(In fact, the Lefschetz Thimble methods can be applied directly to the outer integral in the Sewed Thimble.)
In high dimension, the coordinate of the thimble can only be determined numerically by the flow equation, which is usually a heavy task as the tangent vectors should also be transported simultaneously.
For each point in the sampling, such a flow shall be computed, although only the final coordinate information will be used.
To make a good usage of the whole flow, we can treat the integration along the flow as the element of sampling.
The drawback is that the observables should be calculated along the flow too, therefore no longer as flexible as in sampling on the point approach.
In the following tests of Sewed Thimble with a few observables, we are going to adopt the sampling on the ray approach.
As a first step to resolving this, we select the distribution function $P(\theta)$
\begin{align}
P(\theta)\equiv\int \ud \tau \;e^{-{\rm Re}(\I)+\ln|{\rm Det}J|}.
\end{align}
The original expression (\ref{eq:O_theta}) then corresponds to computing
\begin{align}\label{eq:O_sampled_with_P}
\langle {\mathcal O} \rangle
=\frac{
\Big\langle {\mathcal O}(\theta)/ P(\theta) \Big\rangle_{P(\theta)}
}{
\Big\langle  A(\theta)/ P(\theta) \Big\rangle_{P(\theta)}
},
\end{align}
where the subscript $P(\theta)$ means the sampling is drawn according to the distribution function $P(\theta)$, and
\begin{align}
&A(\theta)\equiv \int \ud \tau \;e^{-i{\rm Im}(\I)+i{\rm Arg}({\rm Det}J)} e^{-{\rm Re}(\I)+\ln|{\rm Det}J|},
\\
&{\mathcal O}(\theta)\equiv \int \ud \tau \;{\mathcal O}(\tau,\theta) e^{-i{\rm Im}(\I)+i{\rm Arg}({\rm Det}J)} e^{-{\rm Re}(\I)+\ln|{\rm Det}J|}.
\end{align}
These definitions are non-trivial. We sample values of $\theta$ using the distribution $P(\theta)$, which arises from integrating over the radial coordinate $\tau$, without the phase and without the observable. We then compute $A(\theta)$ as an integral over $\tau$ with the phase, and  ${\mathcal{O}}(\theta)$ as an integral over $\tau$ including the phase and the observable. These integrals are done numerically. 
\begin{figure}[t]
\begin{tabular}{lr}
\includegraphics[width=0.5\linewidth,trim=0.cm 0cm 0cm 0cm, clip]{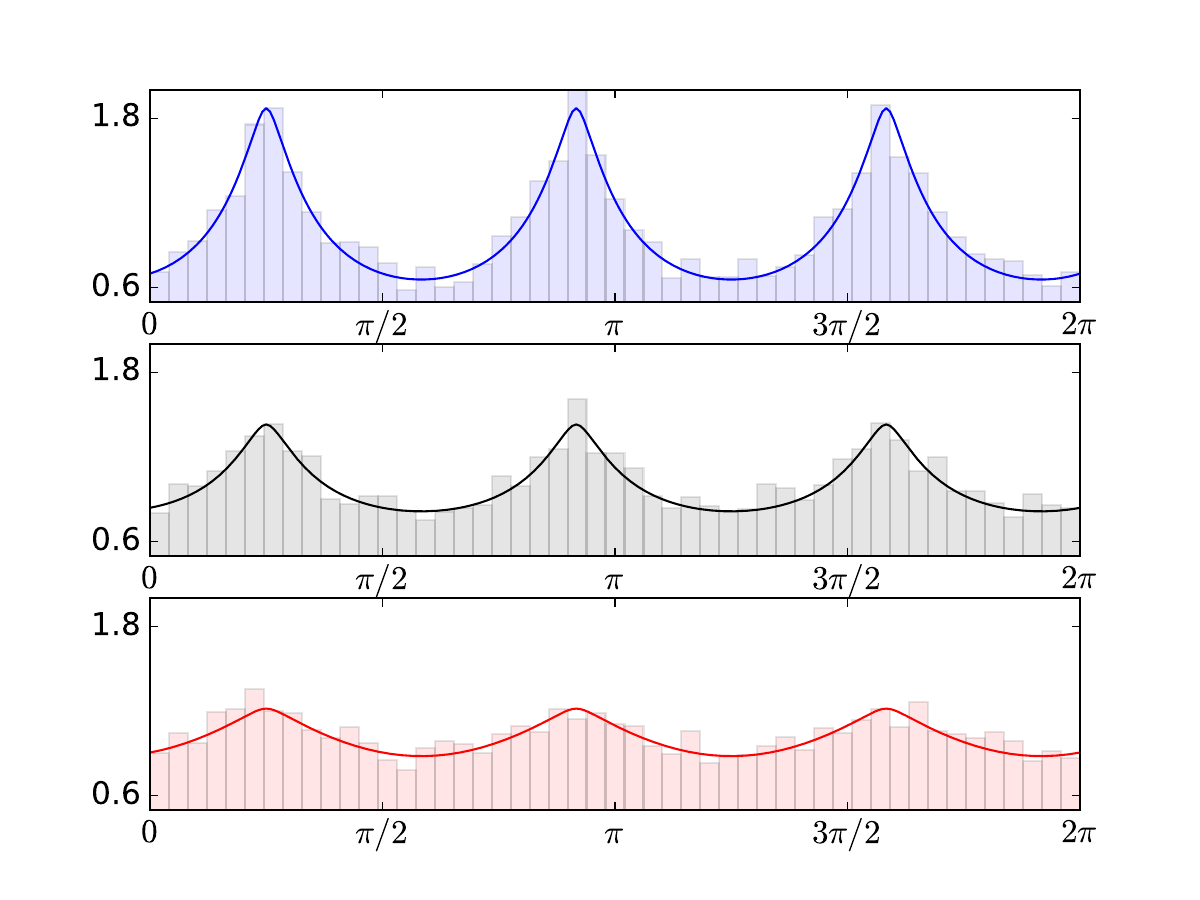} &
\includegraphics[width=0.5\linewidth,trim=0.cm 0cm 0cm 0cm, clip]{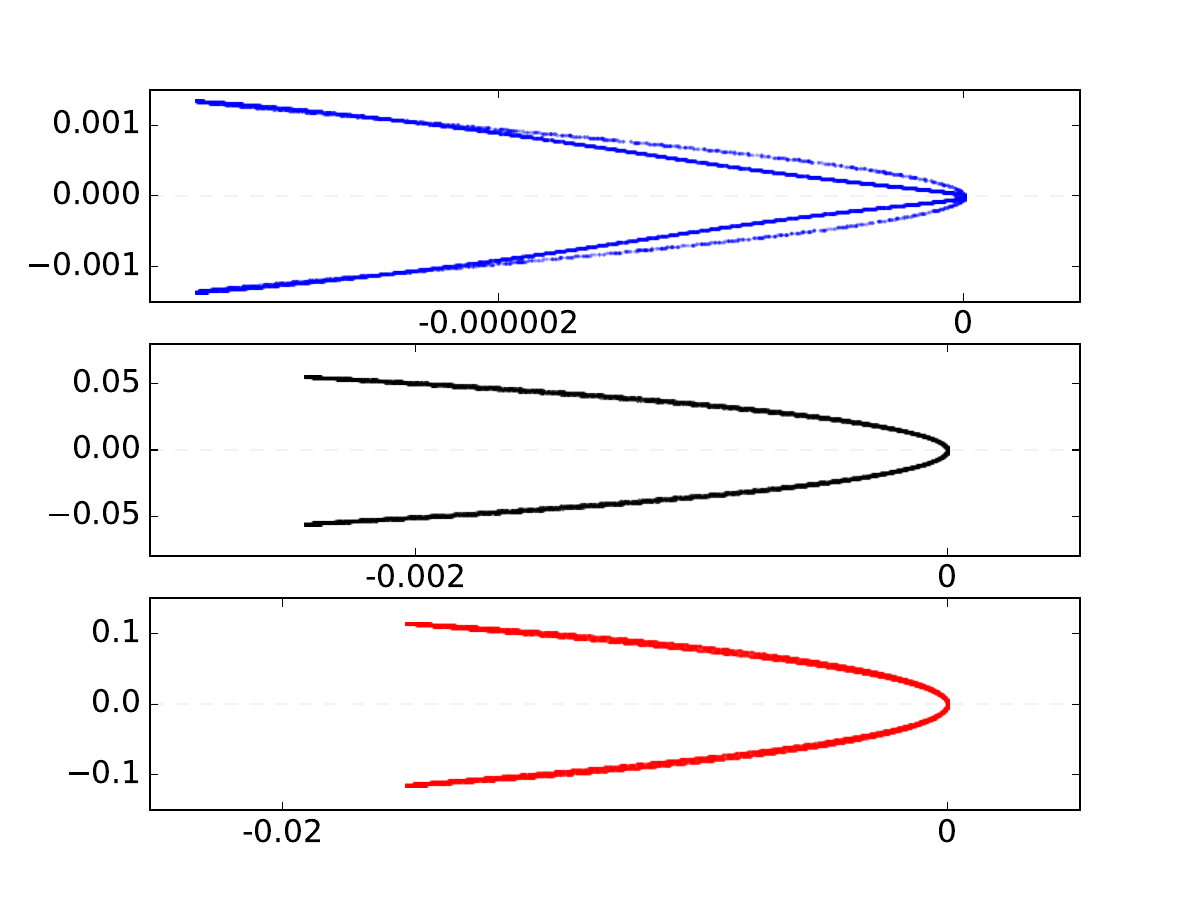}
\end{tabular}
\caption{
(L)
Histogram of $P(\theta)$, where
the solid lines are from Figure \ref{fig:r_theta}
From top to bottom, $r_0=0.1$ (blue), $1$ (black), $2$ (red) respectively.
$p_0=1$,
$a=1$,
$b=1$,
$c=0.1$.
(R) Scatter plot of $A(\theta)/P(\theta)$.
Each dot represents $\left({\rm Re}(A/P)-1, {\rm Im}(A/P)\right)$ of a saved data point.
}
\label{fig:APscatt}
\end{figure}
In Figure \ref{fig:APscatt} (left), we show the histograms of values of $\theta$ for 20,000-step Markov chains, for different values of $r_0$. Overlaid is the analytic weight distribution, which we see matches very well. Apparently at these values of $r_0$, the peak structure does not hamper the MC process. Figure \ref{fig:APscatt} (right) shows the region probed by the complex phase in the denominator $A/P$ of (\ref{eq:O_sampled_with_P}), and we see that it indeed becomes smaller as $r_0$ becomes smaller, and the exponential suppression kicks in closer to the critical point\footnote{The phase coming from the Jacobian $J$ is not suppressed, but since it generically has a power-law dependence on the radius, the exponentially decreasing $e^{-\I}$ will dominate. In some special cases, the Jacobian can become exponential, and even delay convergence. We will return to this point in future.}. We may then proceed to compute the observable $\langle y^4\rangle$, Table \ref{tab:y4MC}, where we discover the numerical calculation reproduces the analytic result, within the 1$\sigma$ error bars.

\begin{table}
\centering
\begin{tabular}{|c|c|}
\hline
$r_0$&$\langle y^4\rangle$\\
\hline
$2$&$3.50\pm 0.10$\\
$1$&$3.37\pm 0.13$\\
$0.1$&$3.34\pm 0.12$\\
\hline
Exact&$3.4$\\
\hline
\end{tabular}
\caption{Numerical and analytic evaluation of $\langle y^4\rangle$. $a=1=b$, $c=0.1$.}
\label{tab:y4MC}
\end{table}

%%%%%%%%%%%%%%%%%%%%%%%
\subsubsection*{Stereographic coordinates:}
%%%%%%%%%%%%%%%%%%%%%%%%%
\begin{figure}[t]
\begin{tabular}{lr}
\includegraphics[width=0.5\linewidth,trim=0.cm 0cm 0cm 0cm, clip]{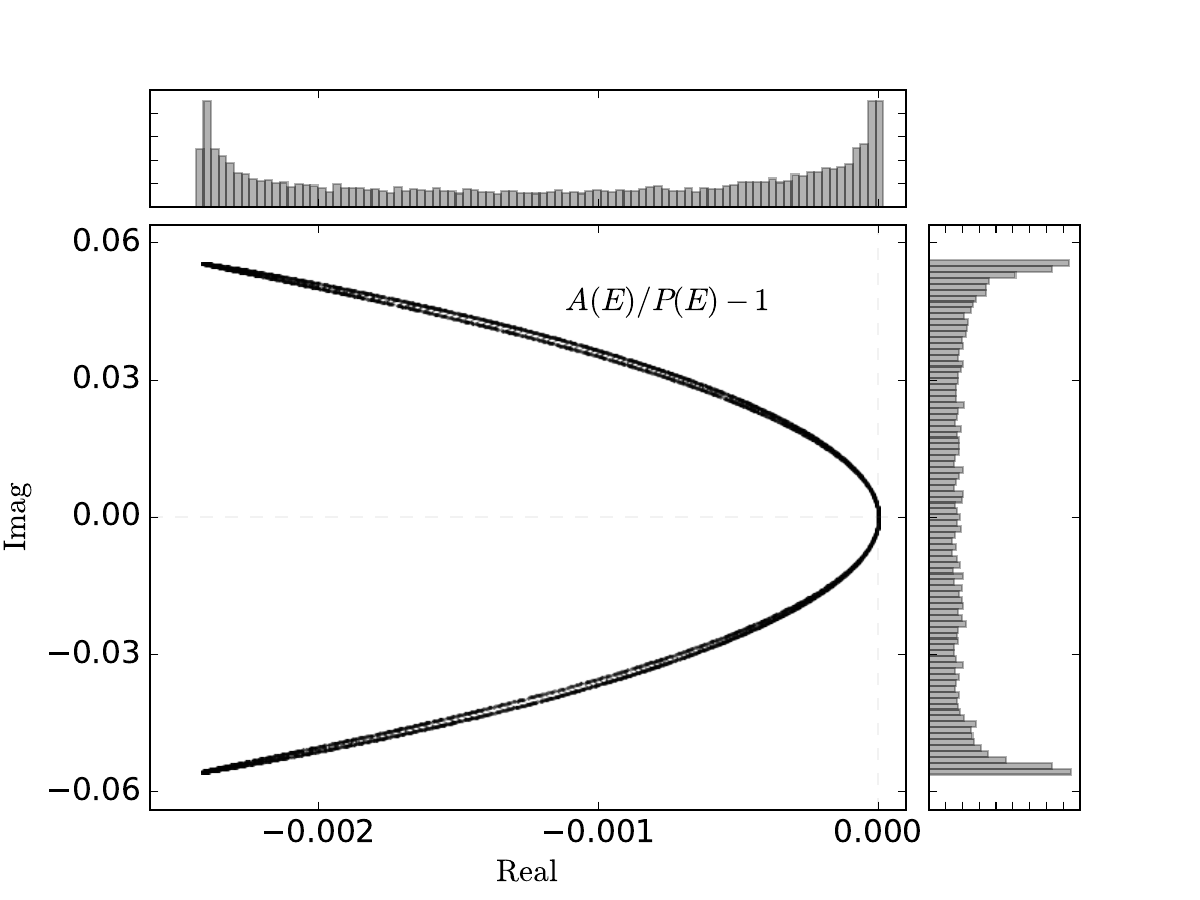} &
\includegraphics[width=0.5\linewidth,trim=0.cm 0cm 0cm 0cm, clip]{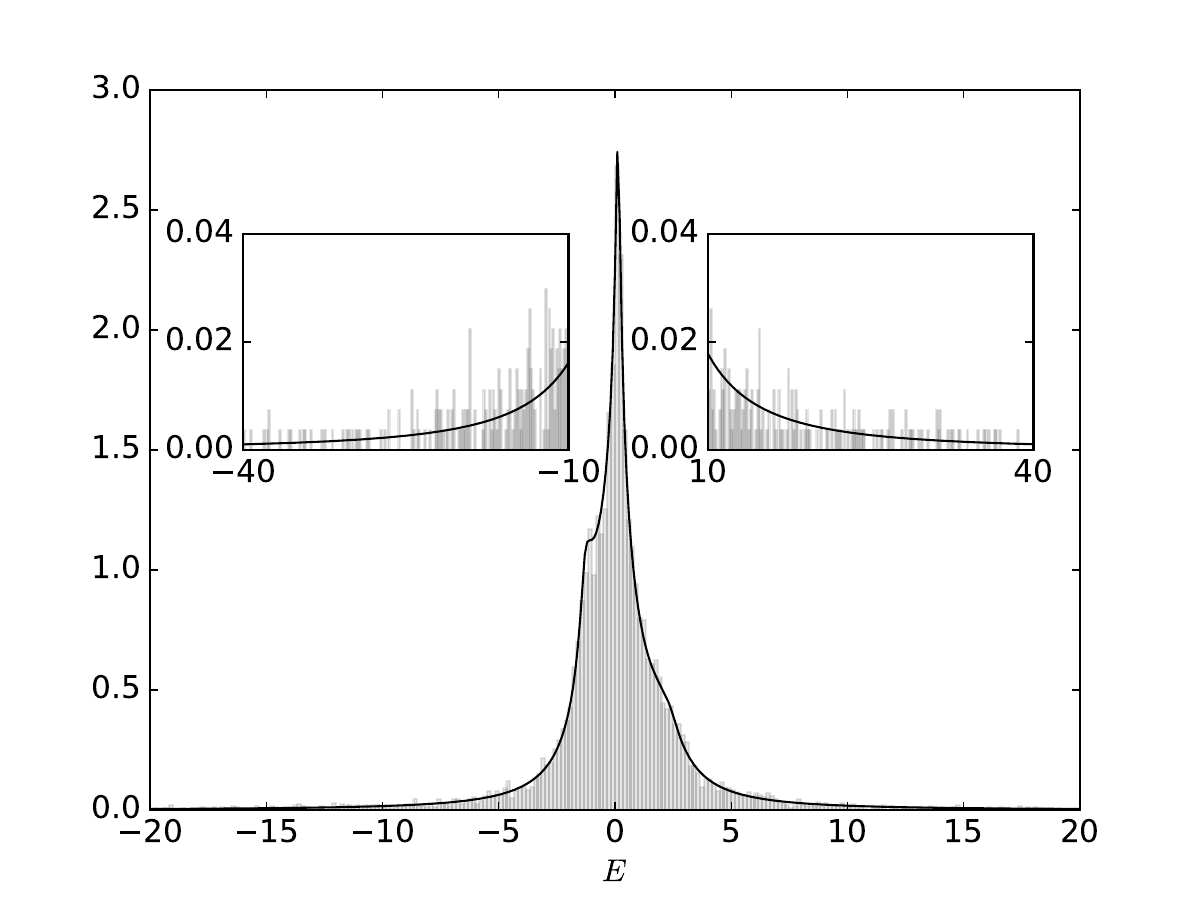}
\end{tabular}
\caption{
(L) Scatter plot of $A(E)/P(E)-1$.
Each dot represents $\left({\rm Re}(A/P)-1, {\rm Im}(A/P)\right)$ of a saved data point.
Histograms on the real and imaginary parts are shown on the sides. 
(R)
Histogram of the probability distribution function $P(E)$.
$r_0=1$,
$p_0=1$,
$a=1$,
$b=1$,
$c=0.1$.
}
\label{fig:Eplot}
\end{figure}
For our example in two variables, we opted for using polar coordinates with one angle $\theta$ and the $\tau$/$r$ radial direction. The finite angular integration interval allows us to compute the integrals along the $\theta$-direction without any compromise on precision.
However, for Monte Carlo sampling of many variables, the periodic boundaries and finite range require extra consideration, which turns out to make this parametrization inefficient.
For this reason, we will parametrize high dimensional integrals using stereographic coordinates.
The switch between the polar coordinate and the stereographic coordinate is straightforward, for a single variable we write (\ref{eq:polar_co-ords}) as
\begin{align}
    e^1&=\frac{2E}{E^2+1},\qquad e^2=\frac{E^2-1}{E^2+1},
\end{align}
with the connection between the two descriptions being $E=\frac{\cos(\theta)}{1-\sin(\theta)}$. Using 100,000 MC steps, we find $\langle y^4\rangle =3.47\pm 0.09$ (for $r_0=1$, see Figure \ref{fig:Eplot}).
Now the weight function has tails for large $E$, since $\ud\theta=\frac{2}{E^2+1}\ud E$, and so is suppressed as $\propto E^{-2}$ as $|E|\to \infty$. The suppression is still only a power-law, and at low dimension (only a few variables), some care is needed for the Markov chain not to get stuck in the tail regions. The situation will be improved in high dimension, as the distribution will go to zero quickly due to the dimensionality.

%%%%%%%%%%%%%%%%%%%%%%%%%%%%%%%%%%%%%%%%%%%%%%%%%%%%%
\section{Application to the discretized real time path integral}
\label{sec:discretized}
%%%%%%%%%%%%%%%%%%%%%%%%%%%%%%%%%%%%%%%%%%%%%%%%%%%%%
\begin{figure}[h]
\centering
\includegraphics[width=0.6\textwidth]{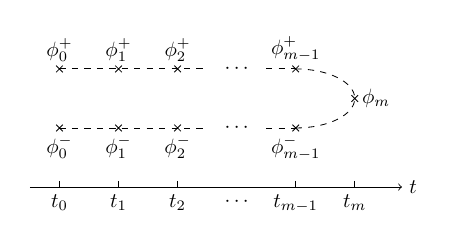}
\caption{Schwinger-Keldysh contour discretized in time.}
\label{fig:SK_contour}
\end{figure}

Ultimately, the purpose of the exercise is to compute expectation values in real-time out-of-equilibrium quantum field theory of the form \cite{Mou:2019tck}
\begin{align}
    \langle\mathcal{O}(t)\rangle&=\mathcal{N}\int\mathcal{D}\phi^+\mathcal{D}\phi^-\mathcal{O}(t)\langle\phi^+_0,\;t_0|\hat\rho|\phi^-_0,\;t_0\rangle e^{\frac{i}{\hbar}\int_\mathcal{C}\mathcal{L}},
\end{align}
where $S=\int_\mathcal{C}\mathcal{L}[\phi]$ is the action and we use the shorthand $\int_\mathcal{C}$ for the combined space and time-integral $\int d^3x\int\, dt$ along the contour $\mathcal{C}$ in the complex $t$-plane (see Figure \ref{fig:SK_contour}). $\mathcal{N}$ is an (infinite) normalisation constant
\begin{align}
    \frac{1}{\mathcal{N}}=\int\mathcal{D}\phi^+\mathcal{D}\phi^-\langle\phi^+_0,\;t_0|\hat\rho|\phi^-_0,\;t_0\rangle e^{\frac{i}{\hbar}\int_\mathcal{C}\mathcal{L}}.
\end{align}
Without further ado, we will rotate to the Keldysh basis \cite{Keldysh:1964ud,Aarts:1997kp,Greiner:1996dx,Kamenev:2009jj}
\begin{align}\label{eq:keldysh_basis}
    \phi^{cl}&=\frac{1}{2}\left( \phi^++\phi^-\right),\quad\phi^q=\phi^+-\phi^-,\\
    \phi^+&=\phi^{cl}+\frac{1}{2}\phi^q,\quad\phi^-=\phi^{cl}-\frac{1}{2}\phi^q,
\end{align}
and specialise to the case of a single real self-interacting scalar field with the potential 
\begin{align}
    V(\phi) = \frac{m^2}{2}\phi^2+\frac{\lambda}{24}\phi^4.
    \label{eq:potential}
\end{align}
 The action may then already at this point be written in terms of field variables on a discrete lattice of space time points as
\begin{align}
    S&=\sum_{x^3}\ud^3 x\,\ud t\left\{\frac{\phi^{cl}_0-\phi^{cl}_1}{dt^2}\phi^q_0-\frac{\phi^{cl}_2-2\phi^{cl}_1+\phi^{cl}_0}{dt^2}\phi^q_1...-\frac{\phi_m-2\phi^{cl}_{m-1}+\phi^{cl}_{m-2}}{dt^2}\phi^q_{m-1}\right.\\\nonumber
    &-\frac{1}{2}\nabla\phi^{cl}_0\nabla\phi^q_0-\nabla\phi^{cl}_1\nabla\phi^q_1...-\nabla\phi^{cl}_{m-1}\nabla\phi^q_{m-1}-\frac{1}{2}V'_0\phi^q_0-\frac{1}{2}\frac{1}{24}(\phi^q_0)^3V'''_0\\\nonumber
    &
    -V'_1\phi^q_1-\frac{1}{24}(\phi^q_1)^3V'''_1
    \left....-\phi^q_{(m-1)}V'_{(m-1)}-\frac{1}{24}(\phi^q_{(m-1)})^3V'''_{(m-1)}\right\}.
\end{align}
where for instance $V_i'=dV/d\phi|\phi_i^{cl}$.

As described in \cite{Mou:2019tck}, we can now split the path integral into two parts, an initial condition and a dynamical part
\begin{align}
    \label{eq:partition_func}
    Z&=\int\mathcal{D}\phi^{cl}_0\mathcal{D}\phi^{cl}_1\;W(\phi^{cl}_0,\phi^{cl}_1) \int\tilde{\mathcal{D}}\phi^{cl}\tilde{\mathcal{D}}\phi^{q}e^{\frac{i}{\hbar}S_{dyn}},\\\label{eq:wigner_func}
    W(\phi^{cl}_0,\phi^{cl}_1)&=\int\mathcal{D}\phi^q_0\langle\phi^+_0,\;t_0|\hat\rho|\phi^-_0,\;t_0\rangle e^{\frac{i}{\hbar}S_{init}},
   \end{align}
   where
   \begin{align}
    \label{eq:S_init_S_dyn}
    S_{init}&=\sum_{x^3}\ud^3 x\,\ud t\left\{ \frac{\phi^{cl}_0-\phi^{cl}_1}{\ud t^2}+\left[\frac{1}{2}\nabla^2\phi^{cl}_0-V'_0\right]\right\}\phi^q_0,\\
    S_{dyn}&=\sum_{x^3}\ud^3 x\,\ud t\sum_{n=1}^{m-1}\left( \left[\frac{-\phi^{cl}_{n+1}+2\phi^{cl}_{n}-\phi^{cl}_{n-1}}{\ud t^2}+\nabla^2\phi^{cl}_{n}-V_n'\right]\phi^q_n-\frac{1}{24}(\phi^q_{(m-1)})^3V'''_{(m-1)}\right),
\end{align}
The object $\tilde{\mathcal{D}}$ does not contain the measures associated with $\phi^q_0$, $\phi^{cl}_0$ and $\phi^{cl}_1$, $S_{init}$ consists of the terms containing $\phi^q_0$, and $S_{dyn}$ is the part of the action with no $\phi^q_0$ terms. This assumes that interactions are switched on at $t=0$, so that there are only linear $\phi^q_0$ terms in $S_{init}$. 

This prescription amounts to stipulating that the initial condition is a free-field (Gaussian) state. The splitting up allows us to sample the initial condition ensemble (the ``initial" integral, variables $\phi^q_0$, $\phi^{cl}_0$ and $\phi^{cl}_1$) directly from a Gaussian initial state with no sign problem. And for each such initial configuration, sample over the remaining variables (the ``dynamical" integral $\phi^q_{>0}$, $\phi^{cl}_{>1}$) on their Sewed thimble as described above. 
The exponent that appears in the Picard-Lefschetz integral is then $\mathcal{I}=-\frac{i}{\hbar}S_{dyn}$,
\begin{align}\label{eq:I_for_PL}
    \mathcal{I}=-\frac{i}{\hbar}\int\ud x\,\ud t\sum_{n=1}^{m-1}\Bigg( &\frac{\overline{\overline{\phi^{cl}_{n+1}}}-\phi^{cl}_{n+1}}{\ud t^2}\phi^q_n-\frac{1}{24}(\phi^q_{(m-1)})^3V'''_{(m-1)}...\nonumber\\&-\frac{1}{(2r-1)!2^{2r-2}}(\phi^q_{(m-1)})^{2r-1}V^{(2r-1)}_{(m-1)}\Bigg),
\end{align}
using the shorthand
\begin{align}
    \overline{\overline{\phi^{cl}_{n+1}}}&=2\phi^{cl}_{n}-\phi^{cl}_{n-1}+\ud t^2\left[\nabla^2\phi^{cl}_{n}-V_n' \right].
\end{align}
The upshot is that once the initial condition variables are chosen, this sets the boundary conditions for the dynamical integral. The critical point is found by solving a second order differential equation $d\I/d\phi_i=0$ (the classical equation of motion corresponding to the action), which has a unique solution given the boundary condition $\phi_0^{cl},\phi_1^{cl}$. This means that for each initial set of variables, there is a unique thimble to sample over.

It remains to define the initial condition through the Wigner function for a free vacuum state
\begin{align}
W\left(\phi_0^{\rm cl},\pi_0^{\rm cl}\right)
=\exp\left(
-\frac{1}{\hbar}
\left[
\omega\left|\phi_0^{\rm cl}\right|^2
+\frac{1}{\omega}\left|\pi_0^{\rm cl}\right|^2
\right]
\right),
\end{align}
with $\omega$ the frequency, and where we have introduced the momentum field variables $\pi_0^{\rm cl}$, which may be chosen through a forward time discretization prescription as
\begin{align}
\pi_0^{\rm cl}\equiv \frac{\phi_1^{\rm cl}-\phi_0^{\rm cl}}{\ud t}.
\end{align}

%%%%%%%%%%%%%%%%%%%%%%%%%%%%%%%%%%%%%%%%%%%%%%%%%%%%%%%%%%%
\subsection{Implementation of the dynamic path integral}
%%%%%%%%%%%%%%%%%%%%%%%%%%%%%%%%%%%%%%%%%%%%%%%%%%%%%%%%%%%

For simplicity, we neglect the spatial directions, so that the system reduces to 0+1 dimensions, quantum mechanics. The dynamical path integral involves a total of $N=2m-2$ field variables. Given values for $\phi_0^{\rm cl}$ and $\phi_1^{\rm cl}$, the unique critical (classical) field variable configuration follows from varying the action $\phi^{\cl}=\tilde\phi^{\cl}$ and $\phi^{\q}=0$.\footnote{Note that the ``classical" variables $\phi^{\cl}$ acquire values corresponding to the classical solution $\tilde\phi^{\cl}$.}

Keeping $\phi_0^{\rm cl}$ and $\phi_1^{\rm cl}$ fixed, the aim is now to rewrite the path integral into the form
\begin{align}
\langle{\mathcal O}(\Phi)\rangle=\int \ud^{N-1} E \int_{0}^{\infty}\ud r\, {\rm Det}(J) e^{-\I}
{\mathcal O}(\Phi),
\end{align}
using stereographic coordinates $E_{1,\ldots,N-1}$ and a radial coordinate $r$, and split the integration domain into an inner (Gaussian flow) and an outer (full non-linear flow) contribution as in the previous section.

Finally, the integrals for an ensemble of initial conditions $\phi_0^{\rm cl}$ and $\phi_1^{\rm cl}$, should be averaged over to produce the value of the chosen observable.

%%%%%%%%%%%%%%%%%%%%%%%%%%%%%%
\subsection{Inner integral}
%%%%%%%%%%%%%%%%%%%%%%%%%%%%%%

The inner integral is computed on the Gaussian thimble, which follows from the Hessian matrix $H$ at the critical point. We want to know the $N$ eigenvalues $\kappa_i$ and eigenvectors $v_i$ according to (\ref{eq:H_eigen})
\begin{align}
H\underline v_\alpha=\kappa_\alpha\underline v_\alpha^{\star}.
\end{align}
We note that $H=iH_{im}$ is purely imaginary, and $H_{im}$ is real and symmetric.
Suppose the eigensystem of $H_{im}$ consists of eigenvalues $s_\alpha$ and eigenvectors $w_\alpha$.
Then, $\kappa_\alpha$ consists of both $s_\alpha$ and $-s_\alpha$, and in particular, for the positive eigenvalues, the eigenvectors can be constructed as
\begin{align}
{\rm if~~} s_\alpha >0 {~~\rm then~~}
&\kappa_\alpha=s_\alpha,~~~v_\alpha=e^{-i\pi/4} w_\alpha,
\\
{\rm if~~} s_\alpha <0 {~~\rm then~~}
&\kappa_\alpha=-s_\alpha,~v_\alpha=e^{i\pi/4}  w_\alpha.
\end{align}
The thimble can then be parameterized as (\ref{eq:z_v_xi})(\ref{eq:xi_sol})
\begin{align}
\underline \Phi=\underline \Phi_{\rm critical}+\sum_{\alpha}c_\alpha r^{\kappa_\alpha}e_\alpha \underline v_\alpha,
\end{align}
where the $c_\alpha$'s are the part relevant to the time shift and can all be set to the same constant, $c_\alpha=c_0$, for simplicity.
The stereographic coordinates $E_i$ are
\begin{align}
\nonumber
e_1&=\frac{2E_1}{\underline{E}^2+1}
,\quad
 \ldots, \quad 
e_{N-1}=\frac{2E_{N-1}}{\underline{E}^2+1}
,\quad
e_{N}=\frac{\underline{E}^2-1}{\underline{E}^2+1}
.
\end{align}
with $\underline{E}^2\equiv\sum_{a=1}^{N-1}E_a^2$.
The Jacobian in the inner region, when converting from $\Phi$ variables in the path integral to $\tau$ and $E_i$, is
\begin{align}
J=\left(\frac{\partial \underline\Phi}{\partial r},\frac{\partial \underline\Phi}{\partial  E_1},\cdots,\frac{\partial \underline\Phi}{\partial  E_{N-1}}\right)
&=\left(\sum_{\alpha }c_\alpha  r^{k_\alpha }e_\alpha \underline v_\alpha \frac{k_\alpha }{r}
,\sum_{\alpha }c_\alpha  r^{k_\alpha }\underline v_\alpha \frac{\partial e_\alpha }{\partial  E_1}
,\cdots
,\sum_{\alpha }c_\alpha  r^{k_\alpha }\underline v_\alpha \frac{\partial e_\alpha }{\partial  E_{N-1}}
\right),
\end{align}
for which the determinant can be written
\begin{align}
{\rm Det}(J_{\rm inner})=
{\rm Det}\left(v\right)
c_0^{N}r^{\sum_{\alpha=1}^N\kappa_\alpha-1}
\frac{2^{N-1}}{(\underline{E}^2+1)^{N+1}}\left(\kappa_N(\underline{E}^2-1)^2+\sum_{a=1}^{N-1}4\kappa_a E_a^2\right).
\end{align}
In the end, the inner integral is calculated as:
\begin{align}
\int \ud^{N-1} E \int_{0}^{r_0}\ud r\, {\rm Det}(J_{\rm inner}) e^{-\I}
{\mathcal O}(\underline \Phi),
\end{align}
where the integration is again limited to the interval $[0,r_0]$ for some choice of $r_0$, and the action $\mathcal{I}$ is the entire non-linear action. We stress that this is not an approximation in terms of the integral that we are interested in, we are just deforming the path away from the thimble, and Cauchy's integral theorem tells us we will get the same answer for the integral.

%%%%%%%%%%%%%%%%%%%%%%%%%%%%%%%
\subsection{Outer integral}
%%%%%%%%%%%%%%%%%%%%%%%%%%%%%%%%

From $\tau_0=\ln(r_0)$ to $r\rightarrow \infty$, the rest of the Sewed thimble is traced out by the gradient flow equations:
\begin{align}
\frac{\partial \Phi_i}{\partial \tau} =\overline{\frac{\partial\I}{\partial \Phi_i}}
,\quad
\frac{\partial}{\partial \tau}\left(\frac{\partial \Phi_i}{\partial  E_a}\right) =\overline{\frac{\partial^2\I}{\partial \Phi_i\partial \Phi_j}\frac{\partial \Phi_j}{\partial E_a}}.
\end{align}
Initialization is set by matching to the Gaussian thimble at $\tau_0$
\begin{align}
\underline \Phi|_0=\underline \Phi_{\rm critical}+\sum_{i}c_\alpha r_0^{\kappa_\alpha}n_\alpha\underline v_\alpha
,\quad
\left.\frac{\partial \underline \Phi}{\partial  E_a}\right|_0=
\sum_{\alpha }c_\alpha r_0^{\kappa_\alpha }\underline v_\alpha \frac{\partial n_\alpha }{\partial E_a}
,\quad \left.\frac{\partial  \Phi_i}{\partial \tau}\right|_0 =\left.\overline{\frac{\partial\I}{\partial  \Phi_i}}\right|_0.
\end{align}
In the end, the outer integral is computed as
\begin{align}
\int \ud^{N-1} E \int_{\tau_0}^{+\infty}\ud \,\tau\, {\rm Det}(J_{\rm outer}) e^{-\I}
{\mathcal O}(\Phi),
\end{align}
where the Jacobian in the outer region is
\begin{align}
J_{\rm outer}\equiv
\left(\frac{\partial\underline\Phi}{\partial \tau},\frac{\partial \underline\Phi}{\partial  E_1},\cdots,\frac{\partial \underline\Phi}{\partial  E_{N-1}}\right).
\end{align}

%%%%%%%%%%%%%%%%%%%%%%%%%%%%%     
\subsection{Reweighting}
%%%%%%%%%%%%%%%%%%%%%%%%%%%%%    

Following the process of section \ref{sec:1variable} we note that the object we wish to compute may be recast as
\begin{align}
\langle {\mathcal O} \rangle
=\frac{
\int \ud  E\, \ud\tau \,{\mathcal O}(\tau, E) e^{-i{\textrm{Im}\mathcal I}+i{\rm Arg}({\rm Det}J)} e^{-\textrm{Re}{\mathcal I}+\ln|{\rm Det}J|}
}{
\int \ud \tau\,\ud E \,e^{-i\textrm{Im}{\mathcal I}+i{\rm Arg}({\rm Det}J)} e^{-\textrm{Re}{\mathcal I}+\ln|{\rm Det}J|}
},
\end{align}
where we have again included the Jacobian as a contribution to the exponent, and split the exponent into a real and imaginary part. The real part gives rise to a positive definite distribution $P(E)$, that we may sample
\begin{align}
\label{eq:Pdistribution}
P( E)=\int \ud \tau\, e^{-\textrm{Re}{\mathcal I}+\ln|{\rm Det}J|}.
\end{align}
While the imaginary part must be included through reweighting. We note the identity
\begin{align}
\label{eq:OAaverage}
\langle {\mathcal O} \rangle
=\frac{
\Big\langle {\mathcal O}( E)/ P( E) \Big\rangle_{P( E)}
}{
\Big\langle  A( E)/ P( E) \Big\rangle_{P( E)}
},
\end{align}
where
\begin{align}
\label{eq:Aintegral}
&A( E)=\int \ud\tau\, e^{i{\rm Arg}({\rm Det}J)-i\textrm{Im}{\mathcal I}} e^{-\textrm{Re}{\mathcal I}+\ln|{\rm Det}J|},
\\
\label{eq:Ointegral}
&{\mathcal O}( E)=\int \ud\tau \,{\mathcal O}(\tau, E) e^{i{\rm Arg}({\rm Det}J)-i\textrm{Im}{\mathcal I}} e^{-\textrm{Im}{\mathcal I}+\ln|{\rm Det}J|}.
\end{align}
The procedure is then to sample $N-1$-tuples of $E_j$ from the distribution $P(E)$ and compute the averages of $A(E)$ and $\mathcal{O}(E)$. Provided the distribution $P(E)$ is sufficiently well-behaved, the integral should converge.

In Appendix \ref{app:A} we present the explicit expressions for the potential (\ref{eq:potential}) for $N=4$ field variables. In the simulations presented below, we use $N=4,6,8$.

%%%%%%%%%%%%%%%%%%%%%%%%%%%%%%%%%%%%%%%%%%%%%%%%%%%%%
\subsection{Monte-Carlo sampling on the thimble}
\label{sec:parameters}
%%%%%%%%%%%%%%%%%%%%%%%%%%%%%%%%%%%%%%%%%%%%%%%%%%%%%

\begin{figure}[t]
\begin{tabular}{lr}
\includegraphics[width=0.5\linewidth,trim=0.cm 0cm 0cm 0cm, clip]{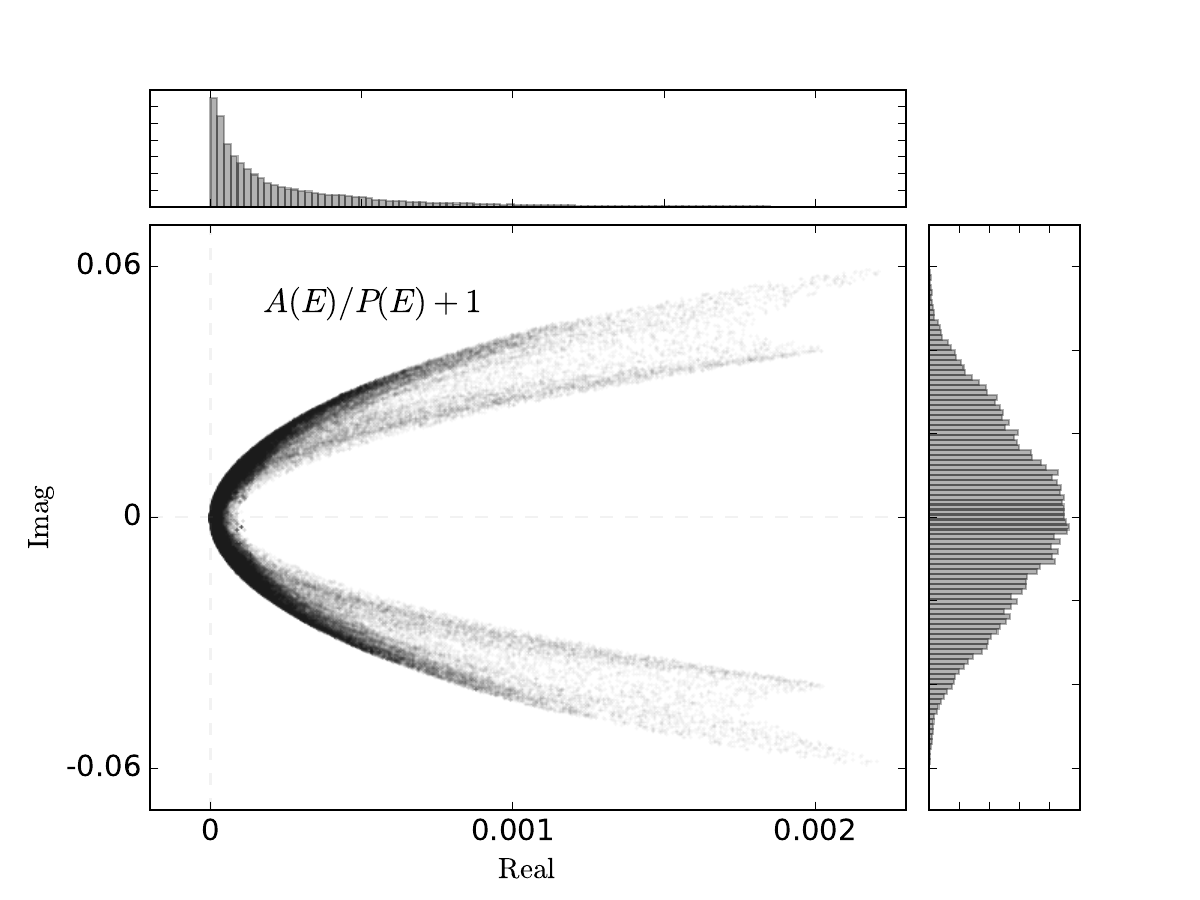} &
\includegraphics[width=0.5\linewidth,trim=0.cm 0cm 0cm 0cm, clip]{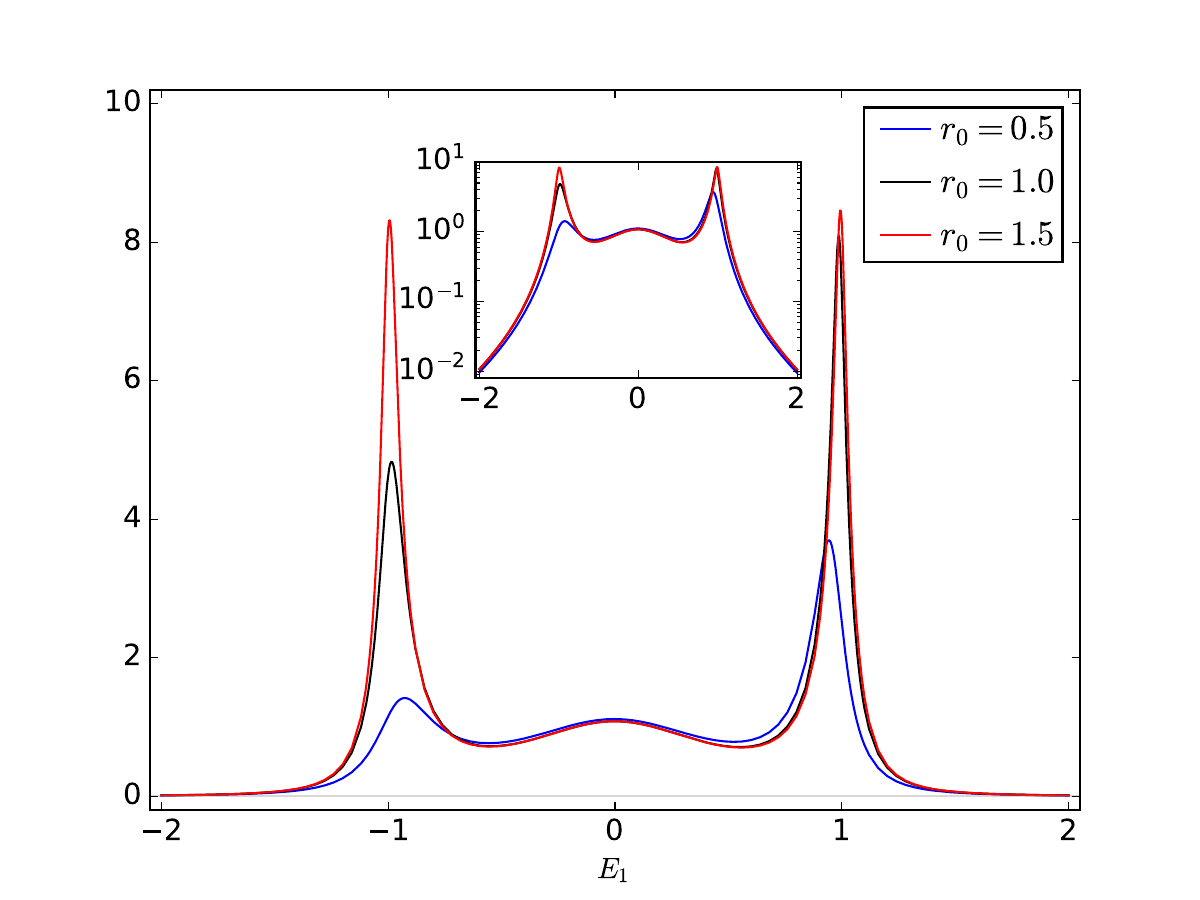}
\end{tabular}
\caption{$N=4$.
(L) Scatter plot of $A(E)/P(E)+1$.
(R)
A scan of the probability distribution function $P(E)$ over $E_1$, with $E_2=E_3=0$.
}
\label{fig:N4dist}
\end{figure}

Just as for the 2-variable model in section \ref{sec:1variable}, we proceed to compute the expectation values in (\ref{eq:OAaverage}) over the distribution (\ref{eq:Pdistribution}) using standard Monte-Carlo techniques. This distribution is again defined through an integral over $\tau$, so as the Markov chain moves around the $N-1$ dimensional $E_a$-space, when performing the Metropolis step, two such integrals must be computed. Computing the observables $A(E)$ and $\mathcal{O}(E)$ also involves performing integrals over $\tau$ given an $N-1$-tuple of $E_a$. All of these involve inner (on the Gaussian thimble up to $r_0$) and outer (on the non-linear thimble from $r_0$) integrals, and we again find discontinuities and distributions similar to Figure \ref{fig:r_theta}. Again, these are harmless as we perform the inner and outer integrals independently.

We have a certain amount of freedom in tuning our implementation. The physical system is determined by the parameters $m^2$ and $\lambda$. The lattice implementation by the parameters $dt$ and $N$. And we must choose a value for $r_0$ ($\tau_0$) as well as parameters of the MC algorithm (proposal function, number of CM steps).

Figure \ref{fig:N4dist} shows the distribution of the object $A(E)/P(E)+1$ (left, similar to Figure \ref{fig:Eplot}) for $N=4$, as well as the probability distribution projected to just one coordinate $E_1$. We see that the phase of $A/P$ again is localized, around $-1$ in this case, and that the sampling of $E$ features a peak structure. The peaks become more pronounced as the Gaussian thimble region is increased, $r_0$ increases.

\begin{figure}[t]
\begin{center}
\includegraphics[width=0.45\linewidth]{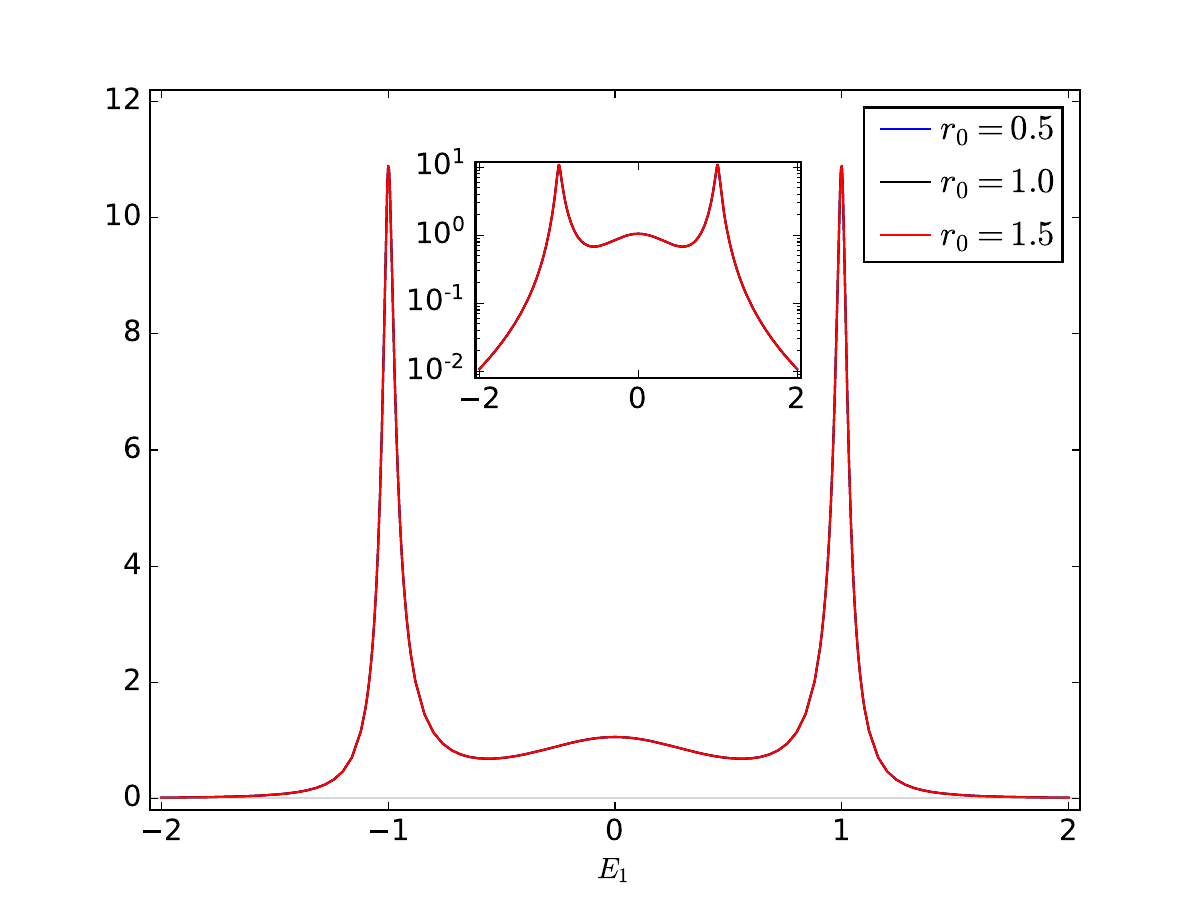} 
\includegraphics[width=0.45\linewidth]{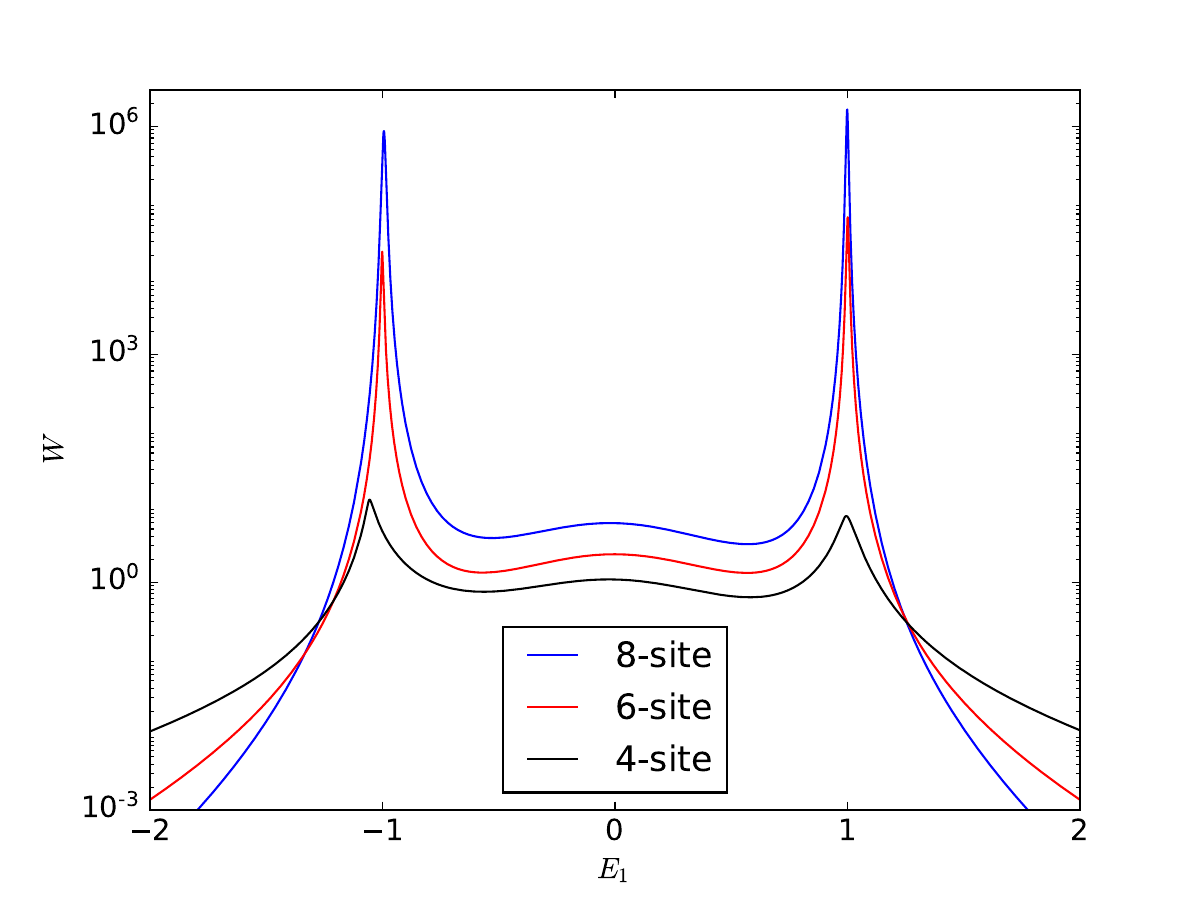} 
\caption{(L) The distribution over $E_1$ for $\lambda =0$, and (R) for different systems sizes $N=4,6,8$, for $r_0=1$.}
\label{fig:distplot}
\end{center}
\end{figure}

In Figure \ref{fig:distplot} (left), we again show the distributions $P(E)$ over just one of the stereographic variables for $N=4$, for different $r_0$ for a non-interacting $\lambda=0$ system. As expected, for $\lambda=0$, there is no dependence on $r_0$, since the whole thimble is Gaussian. In Figure \ref{fig:distplot} (right), we show the distribution for an interacting theory at different systems sizes $N=4,6,8$, clearly showing that the peak structure becomes more pronounced as $N$ increases.

%%%%%%%%%%%%%%%%%%%%%%%%%%%
%\subsection{Results}
%%%%%%%%%%%%%%%%%%%%%%%%%%%
\begin{figure}[t]
\begin{center}
\includegraphics[width=\linewidth]{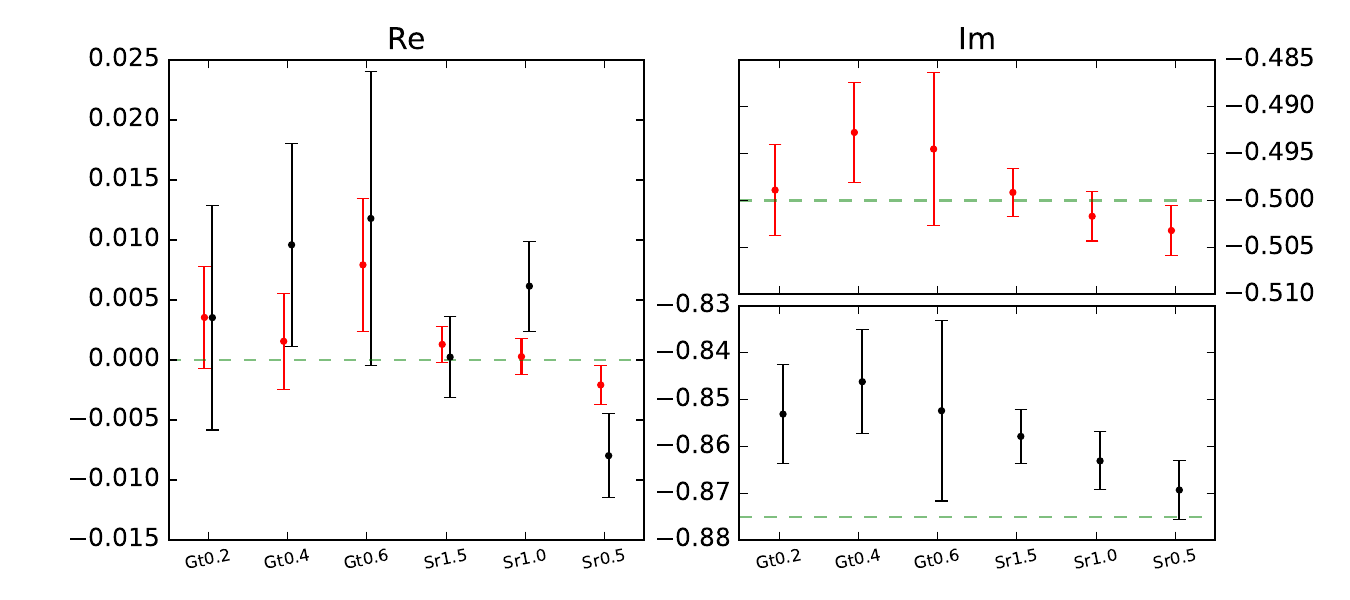} 
\caption{The value of the observables (\ref{eq:obs}) for six different implementations of the dynamic path integral. $\mathcal{O}_2$ is in black and $\mathcal{O}_3$ is in red, with, for example, ``Gt0.2" referring to the Generalized thimble with flow time $\tau_{max}=0.2$ and ``St1.5" referring to the Sewed thimble implementation with $r_0=1.5$. The green dashed lines represent the results of free theory, as analyzed in the Appendix \ref{app:A}.}
\label{fig:operator1}
\end{center}
\end{figure}
We now focus on two observables, the two-point functions
\begin{align}
\mathcal{O}_2=\langle\phi_1^q\phi_2^{cl}\rangle,\qquad
\mathcal{O}_3=\langle\phi_1^q\phi_3^{cl}\rangle,
\label{eq:obs}
\end{align}
and compute them for a single initial condition using six different implementations: The Generalised thimble for flow time $\tau=0.2, 0.4, 0.6$ and the Sewed thimble for $r_0=0.5, 1.0, 1.5$. In all cases, we use $10^6$ MC steps. In Figure \ref{fig:operator1}, We see that they all produce consistent results (the correlators are expected to be purely imaginary), while the Sewed thimble appears to give somewhat smaller statistical errors, a factor of 2-3.

\begin{figure}[t]
\begin{center}
\includegraphics[width=0.6\linewidth]{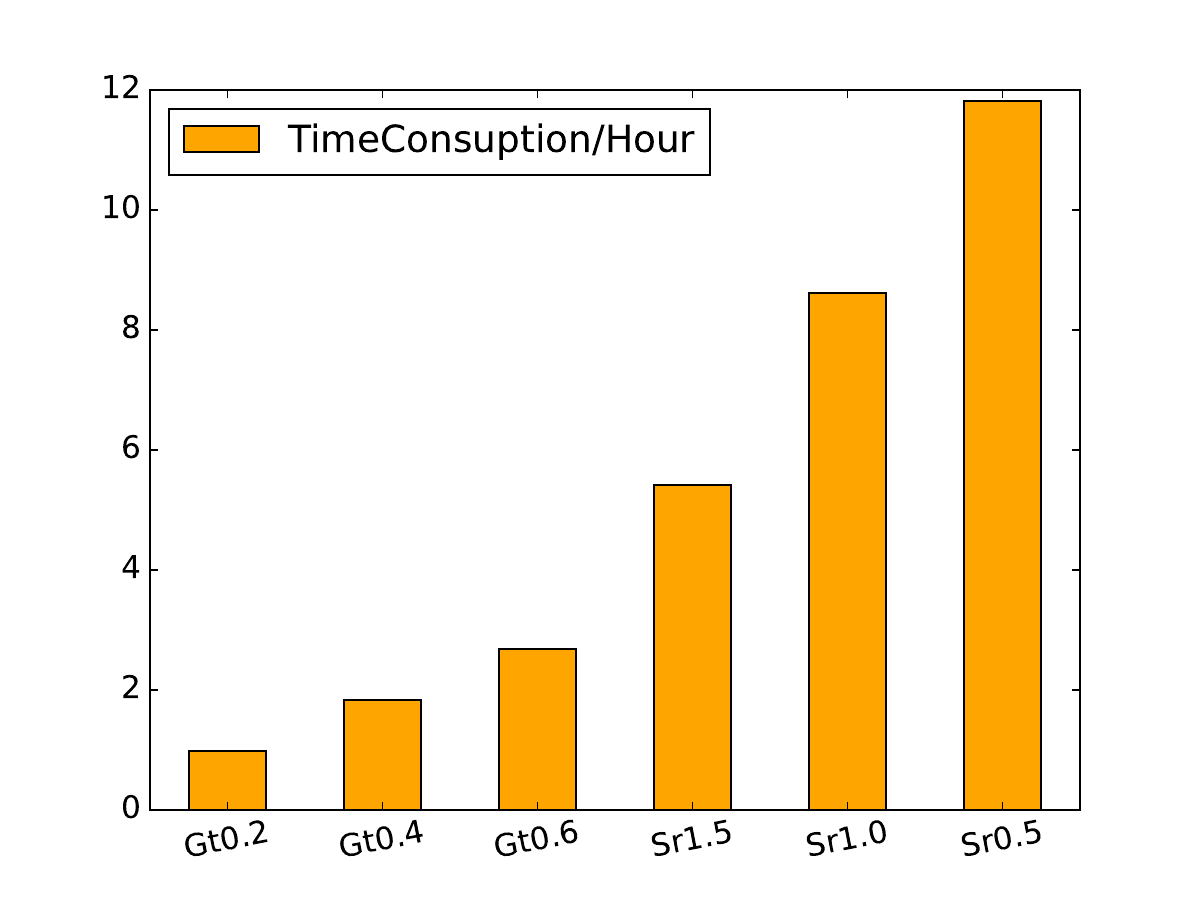}\par
\caption{Wall time for simulations using six different implementations, with the same system size ($N=4$) and the same number of MC steps ($10^6$).}
\label{fig:operator2}
\end{center}
\end{figure}

From a practical point of view the goal is for thimble-MC integration of these path integrals to resolve (or at least ameliorate) the exponential growth of simulation time as the number of variables increases. And so in Figure \ref{fig:operator2} we show the wall-time for the six implementations. We see that for a fixed number of MC steps, the Generalised thimble out-performs the Sewed thimble, and that the choice of algorithm and parameters may give an improvement of up to a factor 10. Combining this with the decrease of statistical errors makes the Generalised and Sewed thimbles fairly evenly matched in terms of performance.

%%%%%%%%%%%%%%%%%%%%%%%%%%%%%%%%%%%%%%%%%%%%%%%%%%%%%
\section{Conclusion}
\label{sec:conclusion}
%%%%%%%%%%%%%%%%%%%%%%%%%%%%%%%%%%%%%%%%%%%%%%%%%%%%%
Resolving the sign problem for real-time quantum fields is a major goal of contemporary numerical field theory. While equilibrium systems at finite chemical potential are coming under control through complex langevin techniques, full quantum real-time evolution out of equilibrium remains a challenge. 

In this work, we have proposed a variant of the method of Lefschetz thimbles, which involves performing MC sampling on a particular type of thimble named ``Sewed" thimbles. These Sewed thimbles consist of a Gaussian inner region near the critical point which can be found analytically in a straightforward way, and an outer region generated numerically by a flow equation. 
By further parameterizing the path integral by a flow time/radius and stereographic variables, a weight function can be generated by integrating over the radius, so that MC sampling need only be done on the remaining $N-1$ variables. 

We find that we can reproduce analytic results for simple systems and results consistent with other thimble approaches, and that in terms of efficiency, Sewed thimbles are competitive with, for instance, Generalised thimbles. 

In this exploratory work, we have been restricted to a very short real-time extent ($N=4,6,8$ variables). But using our stereographic parameterization, this is at least in principle simple to generalise to larger systems. When doing so, one should be aware of a possibly challenging peak structure in the sampling weight function $P(E)$ \cite{DiRenzo:2021kcw}, which for large system may require more sophisticated MC technology than just a Metropolis algorithm. Indeed, multicanonical algorithms or further reparametrizations may be needed to smoothen out these features. 

%%%%%%%%%%%%%%%%%%%%%%%%%%%%%%%%%%%%%%%%%%%%%%%%%%%%%
\begin{acknowledgments}
\label{Sec:acknowledgements}
%%%%%%%%%%%%%%%%%%%%%%%%%%%%%%%%%%%%%%%%%%%%%%%%%%%%%
PMS was supported by STFC consolidated grant number ST/T000732/1. The postdoctoral research of ZM at the University of Southampton is supported by EPSRC (Grant no. EP/W032635/1).
\end{acknowledgments}

%%%%%%%%%%%%%%%%%%%%%%%%%%%%%%%%%%%%%%%%%%%%%%%%%%%%%
\section*{Data Access Statement}
%%%%%%%%%%%%%%%%%%%%%%%%%%%%%%%%%%%%%%%%%%%%%%%%%%%%%

The data on which this paper is based may be reproduced using the code and scripts found on Github at \url{https://github.com/zgmou/SewedThimble}
%\href{https://github.com/zgmou/SewedThimble}{GitHub}.

%%%%%%%%%%%%%%%%%%%%%%%%%%%%%%%%%%%%%%%%%%%%%%%%%%%%%
\appendix
%%%%%%%%%%%%%%%%%%%%%%%%%%%%%%%%%%%%%%%%%%%%%%%%%%%%%

%%%%%%%%%%%%%%%%%%%%%%%%%%%%%%%%%%%%%%
\section{Implementation of the dynamic path integral for $N=4$.}
\label{app:A}
%%%%%%%%%%%%%%%%%%%%%%%%%%%%%%%%%%%%
For completeness, we here provide explicit expressions for quantum mechanics at $N=4$. We will adopt the abbreviations
\begin{align}
& V_0(\phi)   \equiv V(\phi)
,
\quad
& V_1(\phi)  \equiv \frac{\ud V}{\ud \phi}
,
\hspace{3cm}
& V_2(\phi) \equiv \frac{\ud^2 V}{\ud \phi^2}
,
\\
& W_0  \equiv \sum_{l=3,5\cdots} \frac{\phi_1^l}{l!} V^{(l)}(\tilde{\phi}_1^{\rm cl})
,
\quad
& W_1 \equiv \sum_{l=3,5\cdots} \frac{\phi_1^{l-1}}{(l-1)!} V^{(l)}(\tilde{\phi}_1^{\rm cl})
,
\quad
& W_2\equiv \sum_{l=3,5\cdots} \frac{\phi_1^{l-2}}{(l-2)!} V^{(l)}(\tilde{\phi}_1^{\rm cl})
,
\end{align}
where $V^{(l)}(\tilde{\phi}_1^{\rm cl})\equiv \frac{\ud^n V }{\ud \phi^n}\big|_{\phi=\tilde{\phi}_1^{\rm cl}}$.
We can specify the action, and its first- and second-order derivatives, as
\begin{align}
\I=-\frac{i}{2\ud t}\Bigg[
&4\phi_1\tilde{\phi}_2^{\rm cl}
-2\phi_2\tilde{\phi}_1^{\rm cl}
+2\phi_{\underline{2}}\tilde{\phi}_1^{\rm cl}
-4\ud t^2 W_0
\\
&+\left(\phi_2-\phi_1\right)^2
-\left(\phi_{\underline{2}}+\phi_1\right)^2
+\sum_{j=2}^{s}\left[\left(\phi_{j+1}-\phi_{j}\right)^2
-\left(\phi_{\underline{j}}-\phi_{\underline{j+1}}\right)^2\right]
\\
&
-2\ud t^2\sum_{j=2}^s\left[V_0(\phi_j)-V_0(\phi_{\underline{j}})\right]
\Bigg]
,
\end{align}
where $\phi_1\equiv\frac{\phi_1^{\rm q}}{2}$ and with the underscore
$\underline{j}\equiv 2s+2-j$,
\begin{align}
\frac{\partial \I}{\partial \phi}= \frac{i}{\ud t}\left(
\begin{array}{l}
\phi_2+\phi_{\underline{2}}-2\tilde{\phi}_2^{\rm cl}
+2\ud t^2 W_1
\\
\phi_3-2\phi_2+\tilde{\phi}_1^{\rm cl}
+\ud t^2 V_1(\phi_2)
+\phi_1\\
\hspace{0.2cm}\vdots
\\
\phi_{j+1}-2\phi_{j}+\phi_{j-1}
+\ud t^2V_1(\phi_{j})
\\
\hspace{0.2cm}\vdots
\\
\phi_s-\phi_{\underline{s}}\\
\hspace{0.2cm}\vdots
\\
-\phi_{\underline{j+1}}+2\phi_{\underline{j}}-\phi_{\underline{j-1}}
-\ud t^2V_1(\phi_{\underline{j}})
\\
\hspace{0.2cm}\vdots \\
-\phi_{\underline{3}}+2\phi_{\underline{2}}-\tilde{\phi}_1^{\rm cl}
-\ud t^2V_1(\phi_{\underline{2}})
+\phi_1
\end{array} \right)
,
\end{align}
\begin{align}
H\equiv\frac{\partial^2 \I}{\partial \phi^2}=
\frac{i}{\ud t}\left(
\begin{array}{cccccccc}
f_1 & 1  &  &  & & &   &1   \\
1 &f_2  &  1      \\
 & 1  & \ddots   \\
&    & &  & 1   \\
&    & & 1 &  f_{s+1}& -1   \\
&    & &  & -1 & \ddots    \\
&&    & &  &  & &   -1    \\
1&    & &  &  & & -1 & f_{2s}
\end{array} \right)
,
\end{align}
with
\begin{align}
f_1 = 2 \ud t^2W_2
,\quad
f_i =-2+\ud t^2V_2(\phi_i)
,\quad
f_{s+1}=0
,\quad
f_{i} =2-\ud t^2V_2(\phi_{i})
,~~
({\rm if}~i>s+1).
\end{align}

When $s=2$, we have
\begin{align}
\I=-\frac{i}{2\ud t}\Bigg[
&4\phi_1\tilde{\phi}_2^{\rm cl}
-2\phi_2\tilde{\phi}_1^{\rm cl}
+2\phi_{4}\tilde{\phi}_1^{\rm cl}
-4\ud t^2 W_0
\\
&+\left(\phi_2-\phi_1\right)^2
-\left(\phi_{4}+\phi_1\right)^2
+\left(\phi_{3}-\phi_{2}\right)^2
-\left(\phi_{4}-\phi_{3}\right)^2
-2\ud t^2\left[V_0(\phi_2)-V_0(\phi_{4})\right]
\Bigg]
,
\end{align}
\begin{align}
\frac{\partial \I}{\partial \phi}= \frac{i}{\ud t}\left(
\begin{array}{l}
\phi_2+\phi_{4}-2\tilde{\phi}_2^{\rm cl}
+2\ud t^2 W_1
\\
\phi_3-2\phi_2+\tilde{\phi}_1^{\rm cl}
+\ud t^2 V_1(\phi_2)
+\phi_1\\
\phi_2-\phi_{4}\\
-\phi_{3}+2\phi_{4}-\tilde{\phi}_1^{\rm cl}
-\ud t^2V_1(\phi_{4})
+\phi_1
\end{array} \right)
,\quad
H\equiv\frac{\partial^2 \I}{\partial \phi^2}=
\frac{i}{\ud t}\left(
\begin{array}{cccc}
f_1 & 1    &     &1   \\
1 &f_2  &  1      \\
&    1 &  f_{3}& -1   \\
1&  &  -1 & f_{4}
\end{array} \right)
,
\end{align}
where
\begin{align}
f_1 = 2 \ud t^2W_2
,\quad
f_2 =-2+\ud t^2V_2(\phi_2)
,\quad
f_{3}=0
,\quad
f_{4} =2-\ud t^2V_2(\phi_{4})
.
\end{align}

When the interaction is off, the whole integral is a Gaussian one. We can obtain the two-point correlation directly from the Hessian matrix,
\begin{align}
\langle  \phi\phi^T \rangle= H^{-1} .
\end{align}
In the case, we obtain
\begin{align}
&\langle  \phi_1^{q}\phi_2^{cl} \rangle=
-i\ud t,
\\
&\langle  \phi_1^{q}\phi_3^{cl} \rangle=
-i\ud t\left(2-m^2\ud t^2\right),
\end{align}
which are independent of the initial values.

%%%%%%%%%%%%%%%%%%%%%%%%%%%%%%%%%%%%%%
\section{The double integrals}
\label{app:B}
%%%%%%%%%%%%%%%%%%%%%%%%%%%%%%%%%%%%

By performing the integral first on $y$ and then on $x$, we can readily obtain
\begin{align}
\int \ud x\, \ud y\, e^{ -i a  x\left(y-b\right)-icx^3} =\frac{2\pi}{a}
,
\end{align}
where the first integration leads to a delta function, with the definition
\begin{align}
\int \ud p\,  e^{ i pz } =2\pi\delta(z)
.
\end{align}
If instead, we perform the integral on $x$ first, the result is an Airy function, whose definition is
\begin{align}
{\rm Ai}[z]\equiv \frac{1}{2\pi}
\int_{-\infty}^{\infty}\ud t \exp\left(-i\frac{t^3}{3}-itz\right).
\end{align}
This can be used to compute 
\begin{align}
   \int \ud x\, \ud y\, y^4\, e^{ -i a  x\left(y-b\right)-icx^3}
=
\frac{2\pi}{ \sqrt[3]{3c}}
\int \ud y\, y^4{\rm Ai}\left[ \frac{a(y-b)}{ \sqrt[3]{3c}}\right]
=
\frac{2\pi}{a}
\left(b^4+\frac{24bc}{a^3}\right).
\end{align}
Alternatively, we can take use of the delta function
\begin{align}
   &\int \ud x\, \ud y\, y^4 e^{ -i a  x\left(y-b\right)-icx^3}
\\=&
\int \ud y\, y^4\exp\left[-ic \left(\frac{i}{a}\frac{\partial}{\partial y}\right)^3\right]\int \ud x e^{ -i a  x\left(y-b\right)}
\\=&
\frac{2\pi}{a}\int \ud y\, y^4\exp\left[-ic \left(\frac{i}{a}\frac{\partial}{\partial y}\right)^3\right]\delta(y-b)
\\=&
\frac{2\pi}{a}\int \ud y\, \delta(y-b)\exp\left[-ic \left(-\frac{i}{a}\frac{\partial}{\partial y}\right)^3\right]y^4
\\=&
\frac{2\pi}{a}\left(b^4+\frac{24bc}{a^3}\right),
\end{align}
where we have utilized the following convention
\begin{align}
x e^{-iax(y-b)} = \frac{i}{a}\frac{\partial}{\partial y}  e^{-iax(y-b)}
,\quad
\int \ud y f(y) \delta^{(n)}(y-b)=(-1)^{n}\int \ud y f^{(n)}(y) \delta(y-b),
\end{align}
with $f^{(n)}$ denoting the $n$-th order derivative $\partial^n f/\partial y^n$.

%%%%%%%%%%%% references %%%%%%%%%%%%

\end{document}